\documentclass[12pt,preprint]{aastex}

\newcommand{\ms}{$M_{\odot}$}
\newcommand{\msb}{$M_{\odot}$~}
\newcommand{\s}{$s$}

\newcommand{\sbb}{$s$~}

\newcommand{\nq}{$^{14}$N}
\newcommand{\nqb}{$^{14}$N~}
\newcommand{\cd}{$^{12}$C}
\newcommand{\cdb}{$^{12}$C~}
\newcommand{\ct}{$^{13}$C}
\newcommand{\ctb}{$^{13}$C~}

\newcommand{\nvdb}{$^{22}$Ne~}

\newcommand{\nvdanb}{$^{22}$Ne($\alpha$,n)$^{25}$Mg~}

\newcommand{\ctanb}{$^{13}$C($\alpha$,n)$^{16}$O~}
\newcommand{\nqagb}{$^{14}$N($\alpha$,$\gamma$)$^{18}$F($\beta^+\nu$)$^{18}$O~($\alpha$,$\gamma$)$^{22}$Ne~}
\newcommand{\lsim}{\mathrel{\hbox{\rlap{\lower.55ex \hbox {$\sim$}}
 \kern-.3em \raise.4ex \hbox{$<$}}}}
\newcommand{\gsim}{\mathrel{\hbox{\rlap{\lower.55ex \hbox {$\sim$}}
 \kern-.3em \raise.4ex \hbox{$>$}}}}

\usepackage{lscape}     

\shorttitle{$s$-Process Nucleosynthesis}
\shortauthors{Lugaro et al.}

\begin{document}

\title{$s$-Process Nucleosynthesis in AGB Stars: \\
        a Test for Stellar Evolution}

\author{Maria Lugaro\altaffilmark{1}, Falk Herwig\altaffilmark{2},
John C. Lattanzio\altaffilmark{3}, Roberto Gallino\altaffilmark{4},
Oscar Straniero\altaffilmark{5}}

\affil{1. Institute of Astronomy, University of Cambridge,
Madingley Road, \\ Cambridge CB3 0HA, United Kingdom} 
\email{mal@ast.cam.ac.uk}
\affil{2. Department of Physics and Astronomy, University of Victoria, B.C., \\ Box
3055, V8W 3P6, Canada}
\email{fherwig@uvastro.phys.uvic.ca}
\affil{3. School of Mathematical Sciences, PO Box 28M, Monash University, \\ Victoria 
3800 Australia}
\email{j.lattanzio@sci.monash.edu.au}
\affil {4. Dipartimento di Fisica Generale, Universit\'a di Torino, \\ Via Pietro Giuria 
1, 10125 Torino, Italy} 
\email{gallino@ph.unito.it}
\affil {5. Osservatorio Astronomico di Collurania, Teramo, Italy}
\email{straniero@te.astro.it}


\begin{abstract}

We study the $slow$ neutron capture process ($s$ process) in Asymptotic Giant Branch (AGB) stars
using three different stellar evolutionary models computed for a 3 \msb and solar metallicity star.
First we investigate the formation and the efficiency of the main neutron source: the
\ct($\alpha$,n)$^{16}$O reaction that occurs in radiative conditions. A tiny region rich in
\ctb (the \ctb {\it pocket}) is created by proton captures on the abundant \cdb in the top
layers of the He intershell, the zone between the H shell and the He shell. We parametrically
vary the number of protons mixed from the envelope. For high local protons over \cdb number ratio, 
p/\cdb $\gsim$ 0.3, most of the \ctb nuclei produced are further converted by proton capture to \nq. 
Besides, \nqb nuclei represent a major neutron poison. 
We find that a linear relationship exists between the amount of \cdb in
the He intershell and the maximum value of the time-integrated neutron flux. Then we generate
detailed $s$-process calculations on the basis of stellar evolutionary models constructed with
three different codes, all of them self-consistently finding the third dredge up, although with different 
efficiency. One of the
codes includes a mechanism at each convective boundary that simulates time-dependent hydrodynamic
overshoot. This mechanism depends on a free parameter $f$, and results in partial mixing beyond convective
boundaries, the most efficient third dredge up and the formation of the \ctb pocket. For the other two
codes an identical \ctb pocket is introduced in the post-processing nucleosynthesis calculations. 
The models typically produce enhancements of heavy elements of about two orders of magnitude in the He
intershell and of up to one order of magnitude at the stellar surface, after dilution with the convective
envelope, thus generally reproducing spectroscopic observations.
The results of the cases without overshoot are remarkably similar, pointing out that the
important uncertainty in \s-process predictions is the \ctb pocket and not the intrinsic
differences among different codes when no overshoot
mechanism is included. The code including hydrodynamic overshoot at each convective boundary 
finds that the He intershell convective zone driven by the recurrent thermal instabilities of the He shell
(thermal pulses) penetrates the CO core, producing a He intershell composition near to that
observed in H-deficient central stars of planetary nebulae. As a result of this intershell
dredge up the neutron fluxes have a higher efficiency, both during the interpulse periods and
within thermal pulses. The $s$-element distribution is pushed toward the heavier $s$-process elements
and large abundances of neutron-rich isotopes fed by branching points in the $s$-process path
are produced. Several observational constraints are better matched by the models
without overshoot. Our study need to be extended to different masses
and metallicities and in the space of the free overshoot parameter $f$. 

\end{abstract}

\keywords{nucleosynthesis, abundances --- stars: AGB, evolution}

\section{Introduction}

When helium is exhausted in the center of low- to intermediate-mass
stars ($M \lsim$ 8 \ms) the core becomes degenerate \citep{paczynskia} and burning
processes do not occur there anymore. Instead, H and He burn alternately in shells while the star
ascends the Asymptotic Giant Branch (AGB). In this phase the
stellar structure consists of, from the center outward, a degenerate CO core,
the He-burning shell, a thin (10$^{-2} - 10^{-3}$ \ms) zone between the H shell and
the He shell (hereafter He intershell), the H-burning shell, and a large convective envelope, which
suffers from strong stellar winds. When the envelope mass is reduced to less than $\sim 10^{-3}$ \msb 
the star leaves the AGB, a Planetary Nebula is formed and the CO core cools towards the White Dwarf
phase. In 10-20\% of cases the residual H-rich envelope is completely removed by processes associated
with a late He-shell instability during the pre-White Dwarf phase \citep{herwigbld}, leading to
the formation of the H-deficient central stars of planetary nebulae. 

During the AGB the H shell dominates the energy production for most of the time. Hydrogen is transformed into
He, the He intershell consequently grows and is progressively compressed and heated until the temperature and
density become high enough that He burning is triggered in the bottom layers.
The thermonuclear runaway, also known as thermal instability, or {\it thermal  
pulse}, \citep{schwarzschild} generated by this sudden release
of energy causes almost all the He intershell to become convective (we refer to this as the {\it pulse driven
convective zone}, PDCZ) while the envelope expands and the H shell cools. 
Within the PDCZ partial He burning produces large amounts of carbon. The PDCZ 
quenches after a time of a few ten to a few hundred years, while He burning continues radiatively for another 
few thousand years, during which the H shell is inactive. Then, the envelope contracts and shell H burning 
starts again. 
The cycle is repeated for a few up to possibly 100 times, the total number of thermal pulse  depending
on the initial stellar mass and the mass loss rate, 
with thermal periods of a few 10$^{3} up to 10^{5}$ yr.
Recent models of the AGB phase have been computed among others by 
\citet*{hollowell,boothroyda,boothroydb,boothroydc,boothroydd,lattanzioa,
lattanziob,lattanzioc,vassiliadis,blocker,forestini,stranieroc,wagenhuber,mazzitelli,mowlavi,herwig}.

As illustrated in Fig.\ \ref{fig1}, after a limited number of thermal pulses, when the mass of the H-exhausted core is
above $\sim 0.6$ \ms, the convective envelope penetrates in the top layers of the He intershell bringing to
the surface newly synthesized He, C and elements produced by neutron captures. This recurrent phenomenon is
called {\it third dredge up} (TDU). The TDU is responsible for the carbon enrichment shown by MS, S and C(N)
stars. Unfortunately, the reproducibility of the TDU phenomenon is a main problem with AGB computations,
connected to
the more general astrophysical problem of the treatment of turbulent convection in stellar interiors.
Different evolutionary codes do not reproduce the same results:   
some codes found a large amount of TDU at solar metallicity only for intermediate-mass stars 
\citep[5 \msb $\lsim M \lsim$ 8 \ms,][]{ibena}, others found TDU to
occur in low-mass stars ($M <$ 5 \ms), but only for low metallicities: 1/200 of solar 
\citep{ibenrenzinia} and 1/20 of solar \citep{boothroydd}.
\citet{lattanzioc} and \citet{stranieroc} found TDU to occur also in stars
with initial mass as low as 1.5 \msb and solar metallicity. The amount of TDU is also connected to the evolution
of the envelope depending, for example, on the temperature at its base \citep{wood} and its total mass 
\citep{stranieroc}. The inclusion of extra mixing in the computation of TDU, however, can affect these results.
The lower limit of 1.5 \msb at solar metallicity for TDU to occur is consistent with inferences
derived from observational constraints of carbon stars \citep*{groenewegenvd}. \citet{frost} studied the
effect of the numerical treatment on the occurrence and magnitude of TDU, while 
\citet{mowlavi} tested the effect of using an extra mixing mechanism at the base of the
convective envelope and found that TDU occurs in his models only if this procedure is applied.
Recently, \citet{pols} showed that when the AGB evolution is
computed by solving simultaneously the structure equations and the
diffusion equations for changes in chemical composition the TDU strongly  
depends on numerical choices, such as how the diffusion coefficients are
averaged.

Another debated problem is a satisfactory description of mass loss during the AGB phase. It is believed
that the
main mechanism responsible for mass loss in red giants is radiation pressure on dust grains. 
Detailed calculations in conjunction with new observations will address this problem. 
Most recent works include \citet{wachter} and \citet{olofsson}.
In the past an often employed way to include mass loss has been 
to use \citet{reimers}'s formula, in which the mass loss rate depends on the
luminosity, radius and mass of the star and is proportional to a free
parameter, $\eta$, whose value can vary (typically from $\simeq$ 0.3 to ten). Other
semi-empirical relations for the mass-loss rate have been proposed, in particular by 
\citet{vassiliadis,blocker} and \citet*{arndt}. \citet{marigo} investigated the effects of molecular
opacities on the evolution of AGB stars as the C/O ratio grows from below to above
unity and found that in the carbon-rich models a notable cooling results and hence an early onset
of the {\it superwind} is expected in the framework
of the \citet{vassiliadis}'s formulation of the mass loss rate.

\subsection{The $s$ process}

The He intershell is the site of the $s$ process, the neutron capture 
nucleosynthesis that occurs when neutron number densities $N_{n}$ are of the order of
$10^6 - 10^{11}$ n cm$^{-3}$. In this range of $N_{n}$ when an unstable nucleus is produced along the
$s$-process
chain of neutron captures, it typically decays rather than capture another neutron \citep{burbidge}.
However, for relatively high values of neutron density or temperature several $branchings$
along the $s$-process path are open and neutron-rich nuclei can be produced \citep{ward}. As a
``rule of thumb'' the $s$
process distribution follows the law $\sigma(A) N_s(A) = constant$, where 
$\sigma(A)$ is the neutron capture cross section of nucleus $A$ and $N_s(A)$ its s-process abundance
by number. This rule is only valid locally because of the presence of the neutron magic nuclei
at $N$ = 50, 82, 126. These nuclei have very low neutron capture cross
sections, which make them behave as bottlenecks for the neutron
flux. As a consequence steps in the otherwise $\sigma(A) N_s(A) =
constant$ distribution are found in correspondence to neutron magic nuclei.
The depth of these steps is a function of the total time-integrated neutron exposure $\tau =
\int^t_0 N_n v_T dt$ (where $v_T$ is the thermal velocity), so that different neutron exposures
lead to different $\sigma(A) N_s(A)$ distributions \citep[see][ page 568]{clayton}.
In AGB stars heavy $s$-process elements are produced by the $s$ process and then mixed 
to the stellar surface by TDU where they have been observed \citep[for general references see]
[]{smithlb,wallerstein,wallersteink,bussogl,abiab}.
In particular, AGB stars are responsible for the $s$-process {\it main component}
\citep{kappelerbw,arlandini,travagliogg}, i.e. a significant fraction of the Galactic production of
elements from Sr to Pb, as well as for the {\it
strong component} \citep{claytonr} feeding a large fraction of solar $^{208}$Pb
\citep{gallinoab,goriely,travagliogb}.

The \ct($\alpha$,n)$^{16}$O reaction is activated at low temperatures ($\gsim
0.8 \times 10^{8}$ K), and plays the major role as neutron source in AGB stars
\citep{hollowell,gallinobprr,kappelergb}. However, a higher amount of \ctb than present in the
H-burning ashes
is needed to reproduce the observed enhancements of heavy elements.
To enable the formation of a region rich in \ct, the \ctb {\it pocket}, a few 
protons from the envelope must penetrate into the He intershell and then react with the 
\cdb nuclei abundantly present there. 
The possibility of penetration of protons from the envelope directly into the convective
PDCZ was found by \citet{ibenc} to be severely inhibited by an entropy barrier induced by 
the still active H-burning shell. A favorable location for the mixing of protons is the sharp
H/He discontinuity that is left over after the occurrence of TDU while the H shell is
temporarily extinguished \citep{ibenrenzinib}. Recently, models including
time-dependent overshoot \citep{herwigbs} motivated by hydrodynamic simulations by \citet*{freytag}
\citep*[see also][]{singh}, and models with rotation \citep{langer} have
been able to create a proton rich layer with $M_{pocket} \sim 10^{-5} - 10^{-6}$ \msb at the top of the
He intershell after the end of the TDU. A similar effect was obtained by imposing a velocity
gradient profile beyond the border of the convective envelope \citep{cristallo}. 
Another description of turbulent perturbation of stable layers below convective zones can be
performed in the framework of internal gravity waves \citep{denis}. In summary, the
occurrence of hydrogen penetration at the top of the He intershell seems plausible, but the mass involved
and the resulting \ctb abundance profile are still to be treated as relatively free
parameters, constrained by the rates of proton captures on the carbon isotopes.

The \s-process mechanism in AGB stars is visualized in Fig. \ref{fig1}. After less than a few
thousand years \ctb and \nqb are produced by proton capture on the abundant \cdb in the top layers 
of the He intershell. Subsequently, after about 10,000 yr, the \ctanb
reaction is activated during the interpulse phase in radiative conditions at a relatively low
temperature: $\sim 0.9 \times 10^{8}$ K \citep{stranierog}. 
Before the end of the interpulse period all \ctb has burnt and the pocket has become enriched in \s-processed
material. This $s$-process nucleosynthesis occurs in radiative conditions, hence there is no interaction among
the different layers of the pocket. Each zone can be treated separately and the efficiency of the neutron
flux is a local property that only depends on the initial p/\cdb ratio in that layer. 
The neutron flux lasts typically 20,000 yr and produces locally high neutron exposures, of the
order of 10$^{-1}$ mbarn$^{-1}$. On the other hand, because the timescale is long, the neutron density remains
low, reaching about $10^7$ n cm$^{-3}$. At the end of the interpulse the pocket is engulfed by
the following PDCZ and mixed with ashes from the H-burning shell and material from the previous PDCZ. A large
amount of \nvdb is present in the PDCZ as a product of the chain \nqagb starting on the abundant \nqb from the
H-burning ashes. In the convective zone the \nvdanb reaction can be marginally activated and a 
second neutron flux occurs. This second neutron burst 
is opposite in features to that in the pocket: it occurs on a timescale of a few years and it
produces low neutron exposures (of the order of $10^{-2}$ mbarn$^{-1}$) with a
high-peaked neutron density, up to $10^{10}$ n cm$^{-3}$. This neutron burst does not typically
contribute much to the overall production of $s$ elements, however it affects the final
abundances of isotopes connected to branching points. For an extensive review on 
the $s$ process in AGB stars see \citet*{bussogw}.

We want to analyze the dependence of \s-process model predictions on the stellar
evolutionary code used to compute them. We compare the results of post-processing
models based on three evolutionary sequences computed with three different codes, one of them
including an overshoot mechanism. The sequences are different in the details
of the description of physical processes like mass loss, and of the numerical
solution scheme adopted. To our purposes these computations represent a typical set
of models of the same star constructed by different groups. 
The paper is organized as follows: in \S 2 we describe the stellar evolutionary codes and 
the evolutionary sequences we used to compute the nucleosynthesis. 
In \S 3 we describe the nucleosynthesis codes, the formation and efficiency of
the \ctb neutron source, and the neutron fluxes and final $s$-process distributions
as computed with the different evolutionary sequences.
In \S 4 we compare the final results with a number of observational constraints. In \S 5 we outline 
our conclusions. A preliminary study was presented by \citet{lugaroh}.

\section{The evolutionary codes}

To establish the main effects and dependencies we consider only the case of a star of
initial mass 3 \msb and solar metallicity. The most recent $s$-process studies
\citep{gallinoab,bussogw,goriely} have shown that
there is an important variation in the resulting $s$-process
distribution with stellar metallicity. The \ctb in the pocket is a primary neutron
source having formed from H and the \cdb resulting from He burning, hence for decreasing metallicities
heavier and
heavier elements are produced because there are more neutrons per iron seed. This property has been shown
to have important consequences in the study of the $s$ process at different metallicities \citep{bussogl}
and hence on the galactic chemical evolution of heavy elements \citep{travagliogg}. We will not
discuss this point to a large extent since we only consider models of solar
metallicity. \citet{bussogl} demonstrated that a
spread in the efficiency of the neutron release in the \ctb pocket is necessary to cover the
spectroscopic observations at any given metallicity.
Again, we will not discuss this result in detail because for sake of a simple
comparison we perform $s$-process calculations using a given \ctb pocket, chosen as a relatively free
parameter in the evolutionary sequences without overshoot, or as resulted from the computation
in the evolutionary sequence with overshoot included. 

We consider three different stellar evolutionary codes: the Frascati RAphson Newton Evolutionary
Code \citep[FRANEC,][]{stranieroc}, the Mount Stromlo Stellar Structure
Program \citep[MSSSP,][]{woodz,frost}, and the EVOL code \citep{blocker,herwig}.
Since the codes are independent they contain several differences, e.g. with respect to the numerical
treatment, opacities and nuclear reaction rates. The most important difference to be noted here, however,
is the treatment of convective instabilities and the associated mixing, which
influences the calculation of the TDU phenomenon and the formation of the \ctb pocket.

In the framework of the mixing length theory mixing velocities should drop to
zero at the boundary of a convective layer where the radiative temperature gradient 
$\nabla_{rad}$ equals the adiabatic temperature gradient $\nabla_{ad}$, 
$\nabla_{rad}-\nabla_{ad}=0$, defining the Schwarzschild boundary. In reality, because of inertia,
convective eddies are more likely to have non-zero velocity and to be dragged across the Schwarzschild
boundary. This idea is the basis of the concept of mechanical overshoot.
However, when convection penetrates into a chemically inhomogeneous region a discontinuity in the
opacity forms at the stability border. When the opacity in the convective region is larger 
than the opacity in the radiative region then $\nabla_{rad}-\nabla_{ad}>0$ at the convective
border seen from the radiative zone. In this condition extra mixing is induced by the change of
composition and this causes the border to advance into the radiative zone. A famous example of such
situation is found at the outer border of the convective core during central He-burning: the opacity
increases, due to the conversion of He into C and O, and the convective region grows in mass
\citep*{paczynskia,castellanig}. The same phenomenon is 
encountered during the TDU phases \citep*{becker,castellanic}: the H-rich convective envelope penetrates
into the He-rich intershell layer that has a substantially lower opacity.

To handle this special condition in AGB models the MSSSP code uses a numerical scheme
which may involve mixing of an additional stable mass shell after each structure
iteration: ``Suppose mesh point $j$ is the outer (inner) edge of a formally convective
zone. By evaluating the radiative to adiabatic ratio $\nabla_{rad}/\nabla_{ad}$ at mesh
point $j-1$ ($j+1$), we can extrapolate to find $(\nabla_{rad}/\nabla_{ad})_{extrap}$ at
$j+1$ ($j-1$). If this value is greater than unity, then we incorporate the $j+1$ ($j-1$)
mass shell into the convective zone, even though the shell may have
$\nabla_{rad}<\nabla_{ad}$ when calculated formally. This procedure is executed, at each
convective boundary, each time the abundances are mixed over a convective region.'' 
\citep{lattanzioa}. Note that in the MSSSP code convective regions are mixed
after each structure iteration. 
A further modification to MSSSP was described by \citet{frost} to allow for the
case where the increasing depth of the convective envelope (with each iteration) was followed
by a retreat of the envelope convection, leading to failure to converge. When this occurs 
the iterations are treated as capturing the essential time-dependent response of the structure to the 
induced mixing. Hence the greatest depth of the convective envelope is taken as the mixed region,
which cannot later be unmixed during subsequent iterations for this model. This mixed abundance
is taken as the composition for subsequent iterations, perhaps leading to convergence with a 
shallower {\it convective} zone but with the deeper {\it mixed} region.

No special algorithms are adopted by the FRANEC code version used here to handle the discontinuity
in the chemical composition and the consequent jump of the opacity that forms at the base of the
convective envelope. Since both MSSSP and 
FRANEC assume instantaneous (homogeneous) mixing, neither of them can find the
proton mixing into the He intershell at the end of TDU needed for the formation of a \ctb
pocket. The main difference between the two codes is the amount of TDU mass, which is
higher in the MSSSP case. 

In EVOL a time- and depth-dependent overshoot mechanism is included at the convective borders within
the star, which induces partial mixing beyond the boundary of homogeneously mixed regions.
This scheme mimics mixing due to an exponential decay in space of the convective velocity field into the
stable layers, as observed in hydrodynamic simulations, following the prescription of \citet{freytag} who
performed two-dimensional hydrodynamic simulations of time-dependent
compressible convection in order to study shallow stellar surface convection zones.
A similar behavior of the decay of the convective velocity field outside the convection
zone has been found by \citet{asida} in two-dimensional hydrodynamic simulations
of the oxygen-burning shell in massive stars. 
The scale height of these type of mixing remains a free parameter ($f$) for the convective
boundaries considered here. This type of hydrodynamical overshoot has also been used by
\citet{mazzitelli} and \citet{schlattl}.

\citet{herwig} showed that this algorithm, when applied to thermally pulsing AGB models, 
is relevant to address some major observational properties. Time-dependent hydrodynamic overshoot
is a process that naturally leads to the formation of a $^{13}$C pocket needed for the occurrence
of the $s$ process over a range of $10^{-6} - 10^{-5}$ \ms. It increases the TDU efficiency
and therefore may better account for low-luminosity carbon stars \citep*{herwigbd}.
It leads to intershell dredge up of C-O core
material in the helium intershell. This enhances the C and O
intershell abundances 
as required by models such as those presented by \citet{herwigbld}
that reproduce observations of hot H-deficient post-AGB stars 
of type PG 1159   \citep{dreizler} and central stars of
planetary  nebulae of spectral type [WC] \citep{leuenhagen}.
All these effects depend on the value of
a free overshoot parameter $f$. Here we want to check whether the inclusion of hydrodynamic overshoot
is in contradiction with other observational informations, in particular those connected with the
activation of the neutron sources and the consequent \s-process nucleosynthesis. 
We use a value of $f$=0.016 at every convective boundary, except at the base of the convective
envelope where we use $f$=0.128 for the computation of the \ctb pocket. The original value of 0.016
was chosen to model main sequence core overshoot in order to reproduce the observed width of the
main sequence \citep{schaller}. However, with this choice the mass of
the \ctb pocket and consequently the total mass of \s-process material is too small to
account for observations \citep[as already pointed out by][]{goriely}. 

\citet{cristallo} also find partial mixing of protons below the
base of the convective envelope with the FRANEC code. They achieve this by means of a time-dependent mixing
scheme \citep[see][]{chieffi} with mixing velocities derived in the framework of the mixing
length theory. To handle the $\nabla_{rad}-\nabla_{ad}$ discontinuity condition during TDU 
\citet{cristallo} assume that the mixing velocity drops to zero below the 
Schwarzschild boundary of the convective envelope with an exponential decline in space. This choice induces
both a deeper TDU and the formation of a tail of the internal H profile, thus allowing the
subsequent formation of the \ctb pocket. Also in this case, the efficiency of the TDU and
the extension of the resulting $^{13}$C pocket depend on the efficiency of the assumed
exponential decline of the velocity field, i.e. a free overshoot parameter $f_v$. The main difference with
respect to the scheme adopted by EVOL is that in \citet{cristallo} the overshoot is induced only where there is a
$\nabla_{rad}-\nabla_{ad}$ discontinuity, i.e. at the base of the convective envelope during TDU.
We will also briefly comment on the $s$-process results obtained by using as
inputs the \ctb pocket and the stellar features of the FRANEC stellar model of 3 \msb and solar metallicity with
\citet{reimers}'s parameter $\eta$ = 1.5 and overshoot parameter $f_v$ = 0.1 presented by
\citet{cristallo}. This computation was followed for 17 pulses with TDUs, resulting in a final
envelope mass of 1.3 \ms. Because of the algorithm of overshoot included at the base of the
convective envelope, the dredge-up mass is higher than in the FRANEC standard case, reaching up a total of 0.06
\msb and C/O is above unity in the last 5 pulses. The \ctb pocket was found to form over a mass of about 4$\times
10^{-4}$ \ms. 

Mass loss has also been modeled in different ways: the FRANEC code uses \citet{reimers}'s
prescription with $\eta=1.5$ throughout the whole evolution, the MSSSP evolution has been
computed with \citet{reimers}'s mass loss with $\eta=1$ until the beginning of the AGB phase and
afterwards using the formula proposed by \citet{vassiliadis}, which accounts for a superwind
phase. The EVOL code uses \citet{reimers}'s mass loss with $\eta=1$ from the pre-Main Sequence to the point
where C/O=1 in the envelope. Afterwards it uses the \citet{arndt} rate for
carbon stars.
 
\subsection{The stellar evolutionary sequences}

We computed $s$-process detailed nucleosynthesis for a stellar model of 3 \msb and solar metallicity
making use of the structural, thermodynamical and chemical characteristics obtained from the three
different evolutionary codes described above. The FRANEC sequence ends 
because of convergence problems after 25 thermal pulses with TDU, when TDU is vanishing because of the low envelope mass, 
the MSSSP sequence also stops for convergence problems after 18 pulses with TDUs, with an envelope mass similar to
that of the FRANEC case. For the EVOL sequence the adopted mass loss law results in a low mass loss
rate. After reaching the 13$^{\rm th}$ TDU we stopped the sequence under the assumption that an efficient
superwind phase will remove the remaining envelope within a few thermal pulses.

In Fig. \ref{fig2} we present some quantities related to the evolution of the convective envelope
during the AGB phase as functions of the mass of the H-exhausted core $M_{\rm H}$: the mass dredged up by
each TDU episode, the cumulative mass dredged up by all previous TDU episodes, the mass of the envelope and the
C/O ratio in the envelope. Each point corresponds to a thermal pulse followed by TDU. The dredged-up mass and the envelope
mass are determined by the different choices made in each code described above. Because of the different mixing
algorithm and assumption on overshoot, the amount of dredged up mass grows from FRANEC, which does not include
any treatment of the discontinuity of the adiabatic to radiative gradient ratio at the base of the convective
envelope, to MSSSP, which attempts to find the neutrality in that ratio, to EVOL, which includes an
algorithm for hydrodynamic overshoot. For the three models, TDU starts after the 9$^{\rm th}$, 10$^{\rm th}$
and 1$^{\rm st}$ thermal pulse respectively, when the core mass has reached
$\simeq 0.61 - 0.63$ \ms. The AGB sequence with EVOL starts with a larger core mass because of overshoot during the main sequence. In the 3 \msb
model presented by \citet{herwig} the third dredge up starts at the 4$^{\rm th}$ pulse. The difference is due to the
different value of $f$
applied at the base of the convective envelope. An interesting consequence of the deep TDU experienced by
the EVOL model is that after the first six pulses the core mass does not grow anymore, rather decreases with
time \citep{herwigsb}.

The evolution of the envelope mass is determined by the choice of the mass-loss law. For FRANEC, the
\citet{reimers}'s formula determines a continuous decrease of the envelope mass, for MSSSP, which uses 
the \citet{vassiliadis}'s mass-loss law, the mass of the envelope is kept quite constant until a superwind sets
in during the last two pulses. In EVOL, which uses \citet{reimers}'s for the first 6 pulses and then
\citet{arndt} when the envelope is carbon rich, the
mass loss is not as large as in the FRANEC and MSSSP cases and the envelope mass is about double than in
the other two cases right to the end of the computed evolution. The C/O ratio at each pulse is determined by the
mass dredged up and the envelope mass, and also by the C and O
abundances in the intershell which are both larger in the EVOL case
because of intershell dredge up. All the models become
carbon rich in the envelope after a given number of dredge up events.

In Fig. \ref{fig3} we compare some interesting features of the He intershell as
functions of the core mass: the interpulse period, the extent in mass of the intershell convective
region, the overlap factor $r$ between subsequent pulses and the bottom temperature of the pulse driven convection zone. On this
last quantity depends the activation of the $^{22}$Ne neutron source. 
While in the FRANEC and MSSSP cases the interpulse period
decreases with time, and is very similar to previous AGB models \citep[such as those presented by][]
{boothroydc}, in the EVOL case the interpulse period increases with time: from $\sim$
60,000 yr to $\sim$ 80,000 yr. This difference is a consequence of fact that in EVOL the core mass
decreases with time, since the lower the core mass, the larger the interpulse period \citep{paczynskib}. 
For the three stellar evolutionary codes examined, the mass
of the pulse driven convection zone decreases with time and is higher in the EVOL case because the convection extends
down in the CO core. 
The overlap factor $r$ decreases with time in all cases. The
decrease is steeper for cases with larger TDU. The maximum temperature at the base of the pulse driven convection zone  grows
with the pulse number in all three cases. The FRANEC trends are very close to those of
\citet{boothroydc}. In EVOL the
maximum temperature reached is 3.20 $\times 10^8$ K, with respect to the maximum value of 3.05 $\times
10^8$ K reached in the FRANEC and MSSSP case. This higher temperature is a consequence of the 
hydrodynamical overshoot applied at the base of the intershell convection zone.

\section{The nucleosynthesis calculations}

We have used two different post-processing nucleosynthesis codes to different aims.
To investigate the formation of the \ctb pocket and evaluate the efficiency of the \ctb main neutron source 
we have made use of the Monash nucleosynthesis code; to perform detailed $s$-process
calculations we have used the $s$-process code by  
\citet[][]{gallinoab}.

The Monash nucleosynthesis code reads a file of stellar structure inputs such as temperature and density that
is produced by the MSSSP evolutionary code. The nuclear network is specifically designed for the detailed
study of light-element nucleosynthesis in AGB stars. In the nucleosynthesis network there are 74 nuclear
species: from neutrons and protons up to sulphur there are 59 nuclei, with another 14
iron group species to allow neutron capture on iron seeds. There is also an
additional ``particle'' $g$ which has the function of counting the number of neutron captures occurring beyond
$^{60}$Ni hence giving an estimation of the resulting heavier elements $s$-process distribution. All proton,
$\alpha$, neutron captures and $\beta$ decays involving the species listed above are
included in the nuclear network summing up to 506 reactions. The bulk of reaction
rates are from the Reaclib Data Tables of nuclear reaction rates based on the 1991 updated version of
the compilation by \citet{thielemann}. They have been updated
to the latest new experimental or theoretical estimates \citep{lugaro}. 
The reaction network is terminated by a neutron capture on $^{61}$Ni followed
by an {\it ad hoc} decay with $\lambda$ = 1 s$^{-1}$ producing the particle
represented by the symbol $g$: $^{61}$Ni(n,$\gamma$)$^{62}$Ni $\rightarrow$
$^{61}$Ni + $g$. The Q-value assigned to this reaction is the value of the
reverse reaction $^{62}$Ni($\gamma$,n)$^{61}$Ni. 
Following the method of \citet{jorissen} neutron captures on the missing
nuclides are modeled by neutron $sinks$, meaning that the $^{34}$S(n,$\gamma$)$^{35}$S and the
$^{61}$Ni(n,$\gamma$)$^{62}$Ni reactions are given some averaged cross section values in order
to represent all nuclei from $^{34}$S to $^{55}$Mn and from $^{61}$Ni to $^{209}$Bi
respectively. We obtained $\sigma(^{34}$S)$_{sink}$ = 21.1 mbarn at 8 keV (the typical
temperature of \ctb burning) using a solar distribution for the abundances of nuclei from 
$^{34}$S to $^{55}$Mn. To compute $\sigma(^{61}$Ni)$_{sink}$ we used different abundance
distributions of heavy elements computed in different phases of the $s$ process in AGB stars of
3 \msb and solar metallicity from the models of Gallino et al. (1998). We found that the results do not
vary much: changing by up to 15\% for different distributions. Therefore, we decided to use as typical the
average of such values: $\sigma(^{61}$Ni)$_{sink}$ = 135 mbarn at 8 keV. For stellar models of different
masses or metallicity this parameter might have to be adjusted.
The temporal variation of the neutron density is determined in each layer by the local
equilibrium between neutron production and destruction.

The $s$-process code follows a complete set of neutron captures on 404 nuclei from $^{4}$He up to
Po, where the $s$-process path ends because this element is unstable against $\alpha$
decay. $\alpha$-Capture reactions are considered on light nuclei up to Mg. This code uses structure and
thermodynamic inputs from computed stellar sequences. Reaction rates for neutron production have
been taken, as in the Monash nucleosynthesis code, from \citet{denker} for the \ctanb reaction and from
\citet{kappelerwg} for the \nvdanb reaction, excluding the contribution of the elusive
resonance at 633 keV and using the lower experimental limit for the resonance at 828 keV. Rates for
(n,$\gamma$) reactions are updated to the latest estimates \citep{bao,reifarth,mutti}.
Rates for (n,$\alpha$) and (n,p) reactions are taken from the same sources as in the Monash nucleosynthesis code. Decay
rates are mainly from \citet{takahashi}. 

\subsection{The formation and the evolution of the \ctb pocket}

With the Monash nucleosynthesis code we have studied the formation of the \ctb pocket and the activation of the
\ct($\alpha$,n)$^{16}$O neutron source reaction during a typical interpulse period.
This is done to understand the effect of the $^{14}$N(n,p)$^{14}$C neutron poison reaction and the
subsequent recycling of protons by the $^{12}$C(p,$\gamma$)$^{13}$C reaction, possibly inhibited by the 
$^{13}$C(p,$\gamma$)$^{13}$N reaction and proton captures on other isotopes such as $^{18}$O. 
We perform the calculations with the Monash nucleosynthesis code because it includes proton captures while the $s$-process code
does not. We derive the main general features of the neutron flux in the \ctb pocket and their dependence
on various parameters. 

We generated the \ctb pocket by introducing in the post-processing
Monash nucleosynthesis code computation a certain
number of protons below the H/He discontinuity at the end of the 10$^{\rm
th}$ TDU episode, which follows the 19$^{\rm th}$ thermal pulse in the 3 \msb model computed
with MSSSP. The initial mass fraction of \cdb in the He intershell is $X_{\rm C12}$\footnote{We indicate the
mass fraction $X$ of isotope \cdb with the notation $X_{\rm C12}$, and equivalently for all the other
isotopes.}=0.23, which is a very typical value
for an advanced PDCZ  \citep[see e.g. Fig. 9 of][]{boothroydc}. No \ctb or \nqb from
H-burning ashes are present initially in the He intershell since they have been destroyed by $\alpha$
captures during the previous PDCZ. We performed one-zone model calculations by introducing a constant mass
fraction of protons $X_{\rm p}^{init}$ for each simulation in a region of about $10^{-4}$ \ms.
The typical temperature of the top layer of the He intershell grows from about 2$\times 10^{7}$
K to about 4.5$\times 10^{7}$ K in the $\simeq$ 2000 yr following the end of the TDU. At 4.5 $\times
10^{7}$ K the lifetime of protons with respect to the \cd(p,$\gamma)^{13}$N reaction is $<$ 200 yr, so
that all protons are captured before the onset of H-shell burning, which occurs $\simeq$ 3000 yr after the
end of TDU. Carbon-13 is formed by proton capture on \cdb and \nqb is produced by proton captures on the
newly formed \ct. The number of protons available determines the \ctb and \nqb
final concentrations. For the \cd(p,$\gamma)^{13}$N reaction we used the rate from \citet{caughlan}, for
the rate of the \ct(p,$\gamma$)\nqb reaction we used the new evaluation from \citet{king}, which in this
range of temperature is 1.29 times higher than the previous
evaluation by \citet{caughlan}. 

The results of these computations are presented in Fig. \ref{fig4}. As first shown by \citet{hollowell}
and discussed in detail by \citet{jorissen} and \citet{goriely}, for p/\cdb $<$ 1/10, the number of \ctb nuclei
produced is essentially determined by the number of initial protons. Then, for increasing number of protons an
increasing fraction of the \ctb nuclei suffer a second proton capture and are transmuted to \nq, until, at
p/\cdb $>$ 1, a \nqb pocket emerges next to the \ctb pocket. For $X_{\rm p}^{init} \geq$ 0.02,
p/\cdb $\geq$ 1 and \cdb is strongly depleted, being transformed into \nq. 
Since \nqb is a strong neutron absorber, the region of the pocket which is more efficient for the production of
$s$-process nuclei is where \ctb is more abundant than \nq. 
The results are affected by the choice of the \ct(p,$\gamma$)\nqb rate; when the
lower rate from \citet{caughlan} is used the final amount of \ctb is higher. For example, at
$X_{\rm p}^{init} = $ 0.003, with the older rate we obtain 12\% more \ctb and 21\% less \nqb.

The \ctanb reaction occurs during the interpulse period
because the temperature in the \ct-pocket layers reaches $\sim 10^8$ K before the start
of the next pulse. The lifetime of \ctb with respect to the \ctanb reaction is about
400 yr at this temperature. Thus \ctb is totally destroyed before the onset of the next thermal
pulse. With the MSSSP code the temperature in the \ctb pocket
reaches $10^8$ K after about 43,000 yr (with $\rho \sim 13,000$ g cm$^{-3}$ and 
$X_{\rm He4}=$ 0.72) from the beginning of the interpulse period.
The whole interpulse period lasts for about 52,000 yr. Thus, we concur with \citet{stranierog,stranieroc}
that all the \ctb present in the pocket burns in radiative conditions.
\citep[However, it has to be noted that this might
not occur during the earliest interpulse periods, see e.g. Fig. 4 of][]{herwig}.
The resulting neutron exposure is shown in Fig. \ref{fig4}.
A most important point is the role played by the \nq(n,p)$^{14}$C reaction
and the subsequent partial proton recycling \citep[see also][ and references therein]{gallinoab}.
Nitrogen-14 nuclei indeed act as a strong neutron poison during the \sbb process because of the
relatively high cross sections of the \nq(n,p)$^{14}$C reaction: recent measurements give a Maxwellian
average cross section of 2.04$\pm$0.16 mbarn at 24.5 keV \citep{gledenov} with a very nearly $1/v$ shape.
Neutron capture cross sections of all the other CNO nuclei are typically smaller by about two orders of
magnitude. The protons produced by this reaction are recaptured either by the abundant \cd, and \ctb is produced
again, or by the \ctb itself, which has a proton capture rate about ten times larger than that of \cd. The
overall effect can be considered as a partial recycling of neutrons together with a leaking of the neutron
source. In no case is the proton recycling mechanism efficient enough to leave any \ctb present in the pocket at
the end of interpulse period. When the mass fraction of \nqb is smaller than that of \ct, few neutrons are
captured by \nqb and the $s$-process efficiency is high. In this situation the small recycling effect produces
slightly larger final neutron exposures (e.g. 10\% higher for initial proton number of 0.001) than in the test
case when proton captures on \cdb and \ctb are switched off. When the mass fraction of \nqb is higher than that
of \ct, most neutrons are captured by \nqb to be transmuted into protons.
In this situation the recycling effect produces larger final neutron exposures than if proton captures on
\cdb and \ctb were not included (e.g. 30\% more for initial proton number of 0.02). However, the $s$-process
efficiency is low. For $X_{\rm p}^{init} >$ 0.02 the abundance of \cd, and hence the recycling
effect, is much reduced and the neutron flux is unimportant. This explains
why in Fig. \ref{fig4} the total neutron exposure in the \ctb pocket reaches a maximum and then decreases as a
function of $X_{\rm p}^{init}$. In particular, we found that a maximum $\tau$ of 0.43 mbarn$^{-1}$ can be
achieved, for solar metallicity stars, when $X_{\rm p}^{init}$ is 0.003 \citep[as also
found by][]{goriely}. In view of
this, the highest neutron exposure of 0.44 mbarn$^{-1}$ reached among the three representative  
layers of the \ctb pocket assumed in \citet[ see their Fig. 6]{gallinoab}, has to be
considered as the maximum allowed for stars of solar metallicity when the
mass fraction of \cdb in the He intershell $X_{\rm C12}^{intershell} \simeq$ 0.23. 

The recycling/leaking effect is affected by the choice of the \ct(p,$\gamma$)\nqb rate. When the
lower rate from \citet{caughlan} is used \ctb is less destroyed by proton
captures. For example, at $X_{\rm p}^{init} =$ 0.003, with the lower rate we obtain an increase of
the value of $\tau$ by 18\%. 

As mentioned in the introduction, for AGB stars of metallicity lower than solar the $s$-process efficiency is
higher for the same initial \ctb amount. This is because there are fewer iron nuclei seeds per neutron and
fewer light isotopes to act as neutron poisons. However, as well as \ct, also the \nqb nuclei in the
pocket are of primary origin; i.e. they do not depend on the stellar metallicity but are produced starting
only from the H and He originally present in the star. Hence, the effect of the \nqb neutron poison will
also be important in stars of lower metallicity and detailed computations are needed to address this
point. Also rotation could affect the neutron exposure because in this case mixing continues after
H-burning reignition and \nqb is mixed down into the \ct-rich region during the interpulse period. This
has the effect of strongly lowering the neutron exposure \citep*{herwigll}.

We conducted other simulations in the same 19$^{\rm th}$ interpulse period by 
artificially changing the intershell amount of \cd. As shown in Fig. \ref{fig5} a nearly linear
relationship exists between the final maximum value of $\tau$ in the pocket and the intershell \cdb
because the latter determines the \ctb and \nqb profiles and how many protons are recycled. 
The mass fraction of \cdb in the intershell varies from pulse to pulse \citep[see e.g. Fig. 6 of][ and
Fig. 9 of \citealt{boothroydc}]{schonberner} but it typically reaches a
constant value around 0.20 $-$ 0.25 for the advanced pulses where the TDU occurs in the FRANEC and MSSSP
computations, while $X_{\rm O16}$ is only $\simeq$ 0.005. In the EVOL case instead, when hydrodynamic
overshoot is
included and intershell dredge up is at work, the intershell mass fraction of \cdb varies from 0.30 to about 0.50 and then
decreases towards an asymptotic value around 0.40. The intershell mass fraction of $^{16}$O increases
to about 0.20 and then decreases towards about 0.15 \citep[see Fig. 11 of][]{herwig}. 
The maximum neutron exposure in the pocket will then vary from pulse to pulse following the \cdb abundance. 
   
We have determined here that, given the reaction rates of proton and neutron captures on \cd,
\ctb and \nqb, for a given metallicity and for a given intershell abundance of \cdb there is a maximum
value of the neutron exposure achievable in the pocket, and hence a possible variation of  
neutron exposure is allowed to be in the range where $\tau < \tau_{max}$.  
At solar metallicity the maximum $s$-process efficiency is comparable to the maximum value used by
\citet{bussogl} with FRANEC to explain the spread of neutron exposures required by spectroscopic
observation. 

\subsection{The choice of the \ctb pocket for the $s$-process computations}

To follow the \ctb pocket with the $s$-process code we divided the \ct-rich layer at the 
envelope-core interface into a number of zones of suitable mass and neutron exposure. The nucleosynthesis
in each of these zones is computed individually until the end of the interpulse
period. The resulting final abundances in the pocket are obtained as $X_i^{final} = (\sum
X_i^{zone} M_{zone})/\sum M_{zone}$, where $X_i^{zone}$ is the final abundance of the isotope $i$
in each zone, $M_{zone}$ the mass of each zone, and the sum is done over all the zones. 

To specify the actual \ct-pocket profile to run the $s$-process code with the structure computed by FRANEC and
MSSSP, an initial H profile needs to be chosen which relates the $X_{\rm p}$ x-axis in Fig. \ref{fig4} to
the Lagrangian mass coordinate. The extent in mass of the pocket and of each of its zones has to be considered
as a free parameter in these computations, thus leaving open the question of the specific shape of
the H-profile and the mixing processes that lead to the partial mixing.
We follow the approach of \citet{gallinoab} and take the same profile that is plotted in their Fig. 1. The total
mass for the pocket is $5 \times 10^{-4}$ \ms, only the region of the pocket where the \nqb
abundance is lower than the \ctb abundance, and hence the $s$-process is efficient, is considered and 
more mass is given to the zone characterized by the lower neutron exposures. 
The \ctb profile is kept the same for each cycle, however, as discussed by \citet[ see their
Fig. 6]{gallinoab}, the neutron exposure slightly decreases with the interpulse number. This is because the 
initial composition of the \s-processed material in the \ctb pocket is different due to the interplay of
the different phases through the whole evolution.

For the EVOL case, we used the \ctb pocket self-consistently computed at the
5$^{\rm th}$ interpulse period with the overshoot parameter $f =$ 0.128. 
The initial amount of \cd, \ctb and \nq, together with the resulting final neutron 
exposure as functions of the position in mass are plotted in Fig. \ref{fig6}. 
The total mass of the pocket is $3.7 \times 10^{-5}$ \ms. As noted in the previous
section, the higher the initial \cdb amount, the higher is the maximum $\tau$ allowed
(see Fig. \ref{fig5}). Because of intershell dredge up the intershell mass fraction of \cdb
found with the EVOL code, at this interpulse, is 0.48. The maximum of the neutron exposure is 0.84
mbarn$^{-1}$, double the value obtained with the FRANEC and MSSSP codes, for which $X_{\rm C12}^{intershell}
\simeq 0.23$. We note that this maximum neutron exposure is about 20\% higher than the value of 0.70
mbarn$^{-1}$ obtained with the tests presented in the previous section and shown in Fig. \ref{fig5} 
for the same \cdb abundance in the He intershell. This is due to the fact that the profiles of Fig.
\ref{fig6} have been obtained using the \cd(p,$\gamma)^{13}$N reaction rate from \citet{caughlan}, which,
as already shown in the previous section, produces a small increase in
the neutron exposure. 

\subsection{The neutron flux in the \ctb pocket}

To compare the neutron fluxes in the interpulse period as obtained with the three different evolutionary
sequences we plot in Fig. \ref{fig7} the temperature from the three stellar evolution
calculations for a typical interpulse period and the resulting neutron
density as computed by the $s$-process code in the layer of the pocket characterized by the higher neutron density. 
For FRANEC and MSSSP the temperature shown corresponds to a point in mass at about 3$\times 10^{-4}$ \msb
below the H/He discontinuity left by the third dredge up (TDU), for the EVOL code it corresponds
to the point of the maximum of \ctb in the pocket, which is located at about 
3$\times 10^{-5}$ \msb below the H/He discontinuity left by the TDU.
In all three cases the temperature reaches 10$^8$ K before the onset of the next thermal pulse
so that all \ctb is consumed. In the
FRANEC, MSSSP and EVOL cases a typical interpulse period lasts for $\sim$ 40,000, $\sim$50,000 and
$\sim$70,000 yr respectively (see Fig. \ref{fig3}). The time after the beginning of the interpulse period
at which the temperature reaches 0.8 $\times 10^8$ K is about 10,000 yr for the MSSSP
and the FRANEC codes and 20,000 yr for EVOL. This is because the temperature in the MSSSP and
FRANEC cases has been taken at a deeper, hence hotter, layer than in the EVOL case, however, it will not produce
any important difference in the process of the neutron release. Since all \ctb
is consumed, the total integrated neutron flux depends on the initial amount of \ctb and not on
the way it burns. The neutron flux lasts for about 20,000 yr in the three cases, only being shifted in time
because of the different temperature trends. The neutron density peak reaches
1.5, 0.9 and 2.3 $\times 10^{7}$ n cm$^{-3}$ using the FRANEC, the MSSSP and the EVOL codes respectively.
These
variations are unimportant because the neutron density is too low in any case to affect any branching on the 
\s-process path. Note that in the temperature range of interest, $T \simeq 0.8 - 1 \times 10^8$ K, the
\ctanb rate we use is about 50$-$70\% lower than the one adopted in the NACRE compilation \citep{angulo}.
This difference would not make much change in the overall result. If \ctb burns faster the maximum neutron
density will be somewhat higher, yet still unimportant in affecting the $s$ abundances of branching-dependent
isotopes.

In conclusion, the important difference among the three computations regarding the neutron flux in
the \ctb pocket is the maximum neutron exposure reached within the region. As illustrated in detail in
the previous sections the maximum neutron exposure is doubled in the EVOL case with respect to the FRANEC
and MSSSP cases because intershell dredge up in EVOL produces a double intershell abundance of \cd. In the EVOL computation the
$s$-element distribution will be pushed toward heavy $s$-process elements. 

\subsection{The neutron flux in the convective regions by the $^{22}$Ne neutron source}

We already mentioned in the introduction that $^{22}$Ne is produced in the He intershell
during the pulse driven convection zone (PDCZ) by the \nqagb chain that converts to $^{22}$Ne all the $^{14}$N
significantly present in the ashes of the H burning and ingested in the   
PDCZ. The $^{14}$N in the H ashes derives from the conversion of initial CNO nuclei hence
it depends on the metallicity. However, because primary \cdb (and also $^{16}$O, if intershell dredge up
is at work) is dredged up from the He intershell in the envelope, the abundance of $^{14}$N
in the H ashes also contains a primary contribution that grows with the number and the strength of TDU
episodes \citep{gallinoab}. In Fig. \ref{fig8} we show
the contribution of this primary component to the \nqb ingested in the PDCZs as computed by the
different codes. In our computations the primary \nqb that is present in the \ctb pocket, as
discussed in \S 3.1, does not contribute significantly to the total \nqb ingested by
the PDCZ because the mass of the pocket is very small compared to
the mass of the material processed by the H-burning shell (the typical ratio
$M_{pocket}/\Delta M_{\rm H}$ is around 1/15). 

The $s$-process code activates the \nvdanb reaction when the bottom temperature of
the PDCZ grows above 2.5$\times$10$^8$ K. We took the rate of the \nvdanb
reaction from \citet{kappelerwg}, excluding the elusive resonance at 633
keV and using the lower limit of the strength of the 828 keV resonance.
In the temperature range  of interest, $T \simeq 2-4 \times 10^8$ K, this
rate is within a few percent of that adopted in the NACRE compilation \citep{angulo}, while it is 
about 30\% higher than a most recent determination by \citet{jaeger}. Using this latest estimate would 
have a small effect on the results presented here, only changing little the abundances of the
isotopes connected to the branching points that are very sensitive to the activation of the \nvdanb reaction in
the PDCZ, such as $^{96}$Zr.

To compare the neutron fluxes during the thermal pulse the first important parameter to
consider is the maximum temperature at the bottom of each pulse driven convection zone. As shown in
Fig. \ref{fig3} these temperatures when computed by the FRANEC and MSSSP codes are basically the same.
A maximum temperature of $\simeq 3.05 \times 10^8$ K is found in the last PDCZ. With
the EVOL code higher temperatures are achieved as an effect of the hydrodynamic overshoot applied at the base
of the PDCZ and the maximum reaches $\simeq 3.20 \times 10^8$ K. Hence, the
\nvdanb neutron source works
more efficiently when the computation of the stellar evolution includes overshoot. Not only is the
maximum temperature higher in this case, it is also high for a longer
time and the overall distributions in time of temperature and neutron
density in a PDCZ change. This is illustrated in Fig. \ref{fig9}, where we plot the
bottom temperature trends for the last PDCZ computed by the 
three different stellar evolution calculations, together with
the neutron densities computed by the $s$-process code on the basis of the three evolutionary
codes. The temperature profile in the PDCZ has a narrow peak and hence so does the
neutron density. For the FRANEC and MSSSP computations the neutron exposures 
by the $^{22}$Ne neutron burst are very small, about a 
factor of 20 smaller than those in the \ctb pocket.
Note that this can be different for AGB star models of higher mass or of metallicity lower than solar ($Z
< Z_{\odot}$/2), where the temperatures at the bottom of the PDCZs are higher \citep[see
e.g.][]{lugaroz}.
However, the neutron densities reach up to 10$^{10}$ n cm$^{-3}$ and branching points
are extremely sensitive to this neutron flux \citep{gallinoab}. In the EVOL case the
neutron exposure in the PDCZs reaches almost 0.1 mbarn$^{-1}$, which is a value high enough to
influence the production of the $s$-process elements of the Sr peak. Also the neutron density is higher,
up to about 6 $\times 10^{11}$ n cm$^{-3}$, hence we expect that in the EVOL computation 
neutron-rich isotopes produced by branchings on the $s$-process path will
be produced in greater amounts. An important example is $^{96}$Zr: the branching at the unstable $^{95}$Zr
is open when the neutron density is higher than $\simeq 5 \times 10^8$ n cm$^{-3}$. From Fig. \ref{fig9}, 
we see that this occurs for about a year in the FRANEC and MSSSP cases and for about two years in the case
of the EVOL code. The tail of the temperature distribution is also important in determining some final
abundances of isotopes involved in branchings because of the {\it freeze out} effect \citep*{cosner}. In
any case only a small amount of $^{22}$Ne burns: for the PDCZs presented in
Fig. \ref{fig9}, 2.6\%, 1.1\% and 8.9\% of $^{22}$Ne burns in the FRANEC, MSSSP and EVOL cases,
respectively.

\subsection{The resulting $s$-process distributions}

In Fig. \ref{fig10} we show the production factors with respect to solar abundances from 
\citet{anders} of the $s$-only isotopes in the He intershell material averaged over the dredged up
mass: $X_i^{final}=\sum_{j=1}^N X_i^j M_{TDU}^j/\sum_{j=1}^N M_{TDU}^j$.
We sum over all TDU episodes, $X_i^j$ is the abundance of the isotope $i$
in the He intershell after the corresponding pulse with TDU $j$ and $M_{TDU}^j$ is
the mass dredged up. Also included are the production factors of $^{88}$Sr, $^{138}$Ba and $^{208}$Pb,
representative of neutron magic nuclei at $N$ = 50, 82 and 126 respectively, and of three neutron-rich 
nuclei most
sensitive on branchings: $^{86}$Kr, $^{87}$Rb and $^{96}$Zr. 
The $s$-process abundances approach an asymptotic 
distribution after about 5 to 10 TDU episodes. The production factors in the He intershell are up to a
few hundreds and result in enhancements with respect to Fe of up to an order of
magnitude in the envelope. 

It has been noticed in earlier studies, and pointed out again in recent work 
\citep[see for example][]{bussogw}, that the \s-process abundance distribution is grossly
determined once the neutron capture cross sections and the total number of neutrons available 
are given. In other words the neutron exposure determines the steps in the $s$-process abundance
distribution at neutron magic nuclei and the slope of the distribution between the steps. This means that
the general trends of the distributions shown in Fig. \ref{fig10} depend on the choice
of the neutron exposures in the \ctb pocket.
Although the FRANEC and MSSSP final distributions are very similar
since they use the same \ctb pocket, they show small differences: the more
noticeable of these is that the production factors obtained with the MSSSP code
are slightly lower than those obtained with the FRANEC code; moreover, 
the distributions are slightly different. This can be explained by
the different structure inputs, mainly the mass of the convective regions, which is
higher in the MSSSP case (see Fig. \ref{fig3}) resulting in a higher dilution of the material
from the pocket in the He intershell.  

For the EVOL case the situation is different because of the higher  
neutron exposures in the \ctb pocket that are obtained owing to the large abundance 
of \cdb in the He intershell produced by intershell dredge up. While in the FRANEC and MSSSP cases the isotopes between the
neutron magic nuclei $^{88}$Sr and $^{138}$Ba are more enhanced than the isotopes between the neutron magic
nuclei $^{138}$Ba and $^{208}$Pb, the situation is reversed in the EVOL case because   
of the higher neutron exposures in the \ctb pocket. Also at the very end of the
$s$-process path, $^{208}$Pb and $^{209}$Bi are produced in much greater amounts in the EVOL case. 
As noted in the previous section in the EVOL case the neutron exposure in the PDCZs 
reaches 0.1 mbarn$^{-1}$, about five times more than the maximum value
achieved with FRANEC and MSSSP. In this situation the abundances of isotopes  
with $70 \lsim A \lsim 100$ are mostly produced by the activation of the \nvdb neutron source.
In fact, a simulation of the EVOL case performed by excluding the $^{22}$Ne neutron source 
gave production factors of 2, 15, 30 for $^{70}$Ge, $^{88}$Sr and $^{100}$Ru, instead of 99, 96
and 50, respectively. Furthermore, the abundances of isotopes not affected by branching points from
$^{138}$Ba to $^{204}$Pb resulted to be lower by about 50\%, and the production factor of $^{208}$Pb was
lowered by a factor of 3. The distribution in the region between $100 \lsim A \lsim 138$, which is almost
flat for FRANEC and MSSSP, is instead characterized in the EVOL case by a minimum at the $s$-only
$^{116}$Sn. Such minimum is still present but much less accentuated in the simulation where the $^{22}$Ne
neutron source was excluded.

The different modality of activation of the \nvdb neutron source produces variations in the final
results of isotopes involved in branchings. We include in Fig. \ref{fig10} the production factors of three
interesting isotopes produced by branchings on the $s$-process paths: $^{86}$Kr, which is sensitive to the
branching at $^{85}$Kr; $^{87}$Rb, which is sensitive to the branchings at $^{85}$Kr and $^{86}$Rb; and
$^{96}$Zr, which is sensitive to the branching at $^{95}$Zr. Because of the different neutron density
profiles in the PDCZs (see Fig. \ref{fig9}), $^{86}$Kr, $^{87}$Rb and $^{96}$Zr are produced in much
higher amounts in the EVOL computation. 
The fact that the MSSSP case produces less $^{86}$Kr, $^{87}$Rb and $^{96}$Zr with respect to
the FRANEC case is due to the fact that the MSSSP model has fewer PDCZs with the highest temperature than
the FRANEC because of the effect of the higher mass loss rate adopted. 

Another interesting isotope is the $s$-only $^{152}$Gd whose production depends on the branching
at $^{151}$Sm. This isotope is actually positioned on the neutron-poor side of the
branching and hence it may be surprising that is more produced in the EVOL case which has the
higher neutron density. However, this is explained by the {\it freeze out} effect:
the abundances of isotopes connected with branching points freeze out at the time when
``$\sigma\tau_{rem}\leq 1$, where $\sigma$ is the largest cross section
involved in feeding the $s$-only isotope characterizing the branching, and
$\tau_{rem}$ is the remaining neutron exposure until the end of the pulse''
\citep{kappelergb}. Hence, for higher $\sigma$, such as in the case of the branching at
$^{151}$Sm, final abundances are determined late during the neutron flux when the neutron density is
low. The long tail of the neutron density distribution in the EVOL case (see Fig. \ref{fig9}) allows
enough time for the isotopes connected to the ``high-$\sigma$'' branchings such as $^{152}$Gd to determine
most of their abundances at a late time during the neutron flux.

\section{Comparison with observational constraints}

From the earliest \citep{ulrich,ibenb} to more recent studies 
\citep{arlandini} a major constraint for the $s$ process in AGB stars has been the reproduction
of the solar abundance distribution of the $s$-only nuclei belonging to the main
component (i.e. the $s$-process isotopic distribution for atomic mass number $A > 90$). 
From Fig. \ref{fig10} it is clear that our resulting abundance distributions
are non-solar since the production factors of \s-only isotopes are not all the same.
It has been already pointed out by \citet{gallinoab}, making use of
the FRANEC code, that AGB stars of solar metallicity with our choice of the \ctb
pocket do not reproduce the main component. We can
extend this conclusion to the MSSSP and EVOL cases. However, since the abundance
distribution mostly depends on the neutron exposures achieved in the \ctb pocket, which
in turn depend on the \ctb amount (hence also on the intershell \cdb abundance)
and, as previously discussed, on the stellar metallicity, roughly following the rule $\tau\sim$
\ct$/Z$ \citep{claytonmn}, we can obtain a best fit to the main component by adjusting these
parameters. In the FRANEC and MSSSP cases we need to produce more of the heavier elements beyond the
step at $^{138}$Ba by increasing the neutron exposure by about a factor of two. To accomplish this using 
the same \ctb pocket we have to decrease the stellar metallicity to half solar \citep{gallinoab}. For the
EVOL case, on the other hand, we would need to decrease the neutron exposures
in the \ctb pocket. We could try to accomplish this by using a smaller overshoot parameter at the 
base of the PDCZ, hence having a lower \cdb amount in the intershell, or by including some rotationally
induced smearing, which would mix \nqb in the \ctb pocket during the interpulse thus lowering the neutron
exposure. In the EVOL case, though, we could still face great problems connected
with the activation of the $^{22}$Ne neutron source, such as the overproduction of $^{86}$Kr, $^{87}$Rb,  
$^{96}$Zr, and $^{152}$Gd.

However, we will not pursue this point because ``it is a basic premise of the
\citet{burbidge} theory of nucleosynthesis in stars'' \citep*{claytonf} that solar abundances are expected
to come not from a single star, but rather to be built
up by galactic chemical evolution. The more realistic approach to the
origin of $s$ elements in the solar system has been presented by \citet{travagliogg,
travagliogb} and involves the study of the galactic chemical evolution of heavy elements making use of yields
from AGB stars of different masses and metallicities. Making use of the stellar yields computed with
the FRANEC and $s$-process code, these authors have shown
that AGB stars of low mass can contribute satisfactorily to the Galactic abundances of
the heavy $s$ elements. The roles of the $s$-process
{\it weak component} from massive stars \citep*{lamb,raiteri} and
of neutron capture nucleosynthesis in supernovae of Type II have to be reviewed in order to
fully understand how many realistic contributions to the light $s$ elements up to $A \sim 90$ in the
solar system are to be expected \citep{gallinobls}.

Hence we compare our model predictions with other observational constraints. Such comparison
is not intended to be complete since we are only considering one stellar model of given mass,
metallicity and \ctb pocket. We aim at understanding where the three different
evolutionary sequences stand with respect to each other when related to observational constraints.
The $s$-elements enhancement at the stellar surface for AGB stars of solar metallicity are up to a factor of
ten \citep{bussogl}, which is typically matched by our models after dilution of the He intershell material with
the convective envelope. The ratio [hs/ls]=log[(hs/ls)$_{meas}$/(hs/ls)$_{\odot}]$, where `ls', which is the 
average of light \s-process elements (Y and Zr), and `hs', which is the average of heavy \s-process elements (Ba,
La, Nd and Sm) is a good indicator of the distribution of $s$-elements, and varies in the range $-0.6$ - 0
for AGB stars of solar metallicity \citep{bussogl,abiad}. The trends of our model predictions for [hs/ls] follow
from what we observed in \S 3.5: in the FRANEC and MSSSP cases light \s-process elements are more produced than
heavy \s-process elements and [hs/ls] is negative, while the opposite is true for EVOL. In this case, [hs/ls]
is $\sim$ 0. In principle stars with [hs/ls] $\sim$ 0 cannot be covered by the FRANEC and MSSSP predictions at
solar metallicity, however, they can fit into these model at slightly lower metallicities, hence within
observational uncertainties.

Together with spectroscopic observations of AGB stars we also consider recent high-precision laboratory
measurements of isotopic ratios of \s-process elements in presolar silicon carbide grains \citep[SiC, see review
by][]{hoppe}. The vast majority ($>$90\%) of SiC grains from the Murchison meteorite, the mainstream grains,
most likely condensed in the mass losing envelopes of carbon stars. Note that the
condition C/O$>$1 has to be satisfied for SiC molecules to condense. 
In particular, stars of low mass and around solar metallicity have been identified as the most likely parent
stars of mainstream SiC \citep[see][ for a thorough discussion]{lugaroz}. 
Measurements have been performed on aggregates of grains for Sr, Ba, Nd, Sm and noble gases.
Recently, a new technique, developed at the Fermi Institute of Chicago \citep{nicolussid}, has allowed
measurements of isotopic anomalies of Sr, Zr, Mo and Ba in single SiC of the size of few $\mu$m. 
The ratios $^{88}$Sr/$^{86}$Sr and $^{138}$Ba/$^{136}$Ba are sensitive to the neutron exposure. 
Measurements in aggregates of SiC grains indicates that, on average, the $s$-process distribution produced in
SiC grain parent stars is accounted for by a non-solar distribution produced by neutron fluxes in the
pocket of about half the efficiency needed to reproduce the solar main component, such as that 
obtained in the FRANEC and MSSSP cases. The EVOL case produces instead too high $^{138}$Ba/$^{136}$Ba 
and too low $^{88}$Sr/$^{86}$Sr ratios.
The \ctb pocket generated by the stellar model presented by \citet{cristallo} has a neutron exposure profile with
the same maximum $s$-process efficiency as we used in the FRANEC and MSSSP computations, as $X_{\rm
C12}^{intershell} \sim 0.23$, but with a distribution in mass more weighed towards the regions with higher
$s$-process efficiency. This results in an $s$-process distribution slightly pushed towards heavy $s$ elements,
but little changes in the final results. The main difference is that $^{88}$Sr/$^{86}$Sr ratio 
increases going outside of the range observed in SiC.

All the above considerations, however, depend on the main model uncertainty: the \ctb pocket.
For example, as already mentioned, making use of the FRANEC code, \citet{bussogl} showed that one has to assume a
spread in the \ct-pocket $s$-process efficiency to cover the range of spectroscopic observations at any
metallicity. The same conclusion was reached by \citet{lugarod} when considering single SiC grain data. It would
be plausible to decrease the $s$-process efficiency in the pocket by means of stellar rotation as shown by
detailed modeling \citep{herwigll} so that the effect of rotation may naturally explain the needed spread of
\ct-pocket efficiencies at each metallicity. This effect would also lower the neutron exposure in the pocket
found by EVOL. 

Let us consider a few observational constraints that are sensitive to the 
high neutron exposure deriving from the the activation of the $^{22}$Ne in the PDCZ.
The high efficiency of the $^{22}$Ne neutron source produces in EVOL a situation similar to that 
occurring in intermediate-mass stars and in low-metallicity stars. 
The fact that overshoot at the base of the PDCZ induces a higher occurrence of $\alpha$ captures on
$^{22}$Ne is in contrast with spectroscopic observations of Mg, which have shown that 
Mg isotopes are present in solar proportion in AGB stars of about solar metallicity
\citep*{clegg,smithla}. In the FRANEC and MSSSP cases we have a modest increase of the
$^{25}$Mg/$^{24}$Mg and $^{26}$Mg/$^{24}$Mg ratios in the envelope, from
the solar values of 0.13 and 0.14 respectively, to about 0.16. In the EVOL case these ratios increase more
abruptly, up to values around 0.45 \citep[see also][]{herwig}. 
The [Rb/Sr] ratio in AGB stars is typically negative \citep{lambert,abiab}. 
In the FRANEC and MSSSP cases the [Rb/Sr] ratio is always negative, while in the EVOL case is always positive,
at odd with observations. Large depletions of $^{96}$Zr/$^{94}$Zr are seen in single SiC \citep{nicolussid}
indicating that neutron densities in the stellar sources of these grains must have been low. These measurements
are compatible with the FRANEC and MSSSP cases. The EVOL model produces instead only large positive values of
the $^{96}$Zr/$^{94}$Zr ratio. All these effects depend on the choice of the free parameter $f$ at the base of
the PDCZ.

\section{Conclusions}

We presented detailed computations of the $s$-process abundance distribution of heavy elements
for a 3 \msb star of solar metallicity making use of three different evolutionary codes. One of them
includes a time-dependent hydrodynamic overshoot mechanism and finds the \ctb pocket self-consistently.
Hydrodynamic overshoot mechanism at the base of the convective envelope is a likely process to produce a
\ctb pocket. The alternative mixing process related to rotation may limit the neutron flux by continued
mixing of \nqb into the $s$-process layer during the interpulse phase \citep{herwigll}. A
combination of both processes should be investigated in the future. It is of course conceivable that 
other mixing processes could be involved, which may have complementary or similar properties to those of
the hydrodynamic overshoot mechanism considered here. Overshoot at the base of the convective envelope
generates higher TDU efficiencies making it possible to create carbon stars at very low luminosities. The
inclusion of overshoot at the base of the PDCZ, and the consequent intershell dredge up, help to explain the observed
carbon and oxygen abundances of  H-deficient central stars of
planetary nebulae. 

The two computations that do not include overshoot do not self-consistently find any proton diffusion at
the end of TDU and we assumed an identical \ctb pocket for these cases. The final results of
these two cases are remarkably similar and we can conclude that the main uncertainty in \s-process
predictions is the \ctb pocket and not the intrinsic differences among different codes when no overshoot
mechanism is included. In the case that includes hydrodynamic overshoot, because of penetration of the
intershell convective region into the CO core, (i) the abundance of \cdb in the intershell and hence the
maximum total neutron exposure achieved in the \ctb pocket are higher than in the other cases, and (ii) the
neutron flux in the PDCZs is stronger. In summary, overshoot at the
base of the convective envelope produces a \ctb pocket if large efficiencies, i.e. values of the parameter
$f$, are assumed; overshoot at the base of the PDCZ influences the activation of the
$^{13}$C($\alpha$,n)$^{16}$O and $^{22}$Ne($\alpha$,n)$^{25}$Mg neutron sources producing a more efficient 
neutron flux in both situations. The final resulting abundance distributions of heavy elements reflect
such features of
the neutron fluxes: in the case with overshoot included, the heaviest
elements, such as Sm, are produced in greater amount than lighter elements such as Sn. At the end of
the $s$-process path $^{208}$Pb is produced in a higher amount, as are also isotopes fed by branching points,
such as $^{86}$Kr, $^{87}$Rb and $^{96}$Zr. 

With our choice of the \ct-pocket profile the results from the two cases without overshoot give a
good match to several observational constraints. The strong intershell dredge up connected with overshoot at the base of the
PDCZs is in several situations in contrast with observational constraints. We still need to expand this study to
a range of masses and metallicities. Within the EVOL case, we also need to extend our calculations in the space
of the free overshoot parameter $f$, as applied at different boundaries and possibly also different times during
the evolution, in order to test which range of $f$ could provide us with the closest match to observational
constraints. For example, a preliminary test with $f$=0.008 applied at the base of the PDCZ show a
significantly reduced efficiency in the activation of the $^{22}$Ne neutron source.

\acknowledgements

FH would like to thank D.A.\,VandenBerg for support through his Operating Grant from the
Natural Sciences and Engineering Research Council of Canada.
Work partly supported by the Italian MURST-Cofin 2000 project ``Stellar Observables of Cosmologic
Relevance''.

\clearpage

\centerline{\bf FIGURE CAPTIONS}

\figcaption[fig1]{Structure evolution through two thermal pulses of a 3 \msb star model of solar metallicity
computed with the MSSSP stellar evolutionary code (see \S 2). The y-axis is the Lagrangian mass
coordinate in solar masses. Convective regions are white and it is possible to recognize the pulse driven convection zones during the 
the 19$^{\rm th}$ and 20$^{\rm th}$ pulse, as well as the bottom of the convective envelope.
The grey lines in the radiative regions represent the mass shells. The region with
the higher mesh density is the H-burning shell. The \ctb pocket is a
local phenomenon involving a mass region of the order of half a tick-mark (or less) on the y-axis.
The x-axis is a representation of time by
model number (from the Monash nucleosynthesis post-processing code, see \S 3).
The corresponding approximated time intervals are as follows: starting from t$=$0 at the end of the
19$^{\rm th}$ PDCZ, the 10$^{\rm th}$ TDU lasts for about 160 yr, the formation of the \ctb pocket occurs
within t$=$2500 yr, the \ct($\alpha$,n)$^{16}$O reaction starts at t$=$11,000 yr and by t$=$42,000 yr \ctb
nuclei are depleted to a number lower than 10$^{-6}$. The ingestion of the pocket 
in the PDCZ occurs at t$=$52,000 yr. The PDCZ lasts for about 40 yr, with about 4 yr
during which the bottom temperature is above 2.5 $\times 10^8$ K and the $^{22}$Ne($\alpha$,n)$^{25}$Mg
reaction is marginally activated.\label{fig1}} 

\figcaption[fig2]{TDU mass and envelope features as functions of the mass of the H-exhausted core for the
3 \msb and solar metallicity models computed with the FRANEC (open squares), MSSSP (open triangles) and
EVOL (open circles) codes. Each point corresponds to a thermal pulse followed by TDU. These models were used to
compute the $s$ process. For each sequence the full symbol represents the starting point from
which the evolution develops following the dotted lines connecting the open symbols. 
Note that the core mass always increases with time in the FRANEC and MSSSP cases, while 
in the EVOL case it decreases with time after 6 pulses. For each model we show: {\bf a)} the
mass dredged up by each single TDU episode, {\bf b)} the envelope mass, {\bf c)} the cumulative mass
dredged up, and {\bf d)} the C/O ratio in the envelope.\label{fig2}} 

\figcaption[fig3]{ 
Structural and thermodynamical characteristics as function of the core mass. 
Each point corresponds to a thermal pulse followed by TDU. For each sequence the full
symbol represents the starting point from which the evolution develops following the dotted lines
connecting the open symbols. {\bf a)} 
The interpulse period, note that in the EVOL case the interpulse period increases with the pulse
number, while for FRANEC and MSSSP it decreases. {\bf b)} The mass of the PDCZ, which decreases
with time in all cases. {\bf c)} The overlap factor $r$ between subsequent PDCZs,
which decreases with time in all cases, and {\bf d)} the variation of the maximum temperature at
the bottom of the PDCZs, which increases with time in all cases. 
\label{fig3}}

\figcaption[fig4]{ 
Resulting mass fractions of \cd, \ctb and \nqb as functions of the initial mass fraction of protons
introduced below the H/He discontinuity during the simulations performed with the Monash nucleosynthesis code.
The protons are introduced after the 10$^{\rm th}$ TDU episode of the 3 \msb star of solar
metallicity computed with the MSSSP code. Also plotted is the corresponding neutron exposure at
the end of the interpulse period after all the \ctb has burnt. 
\label{fig4}}

\figcaption[fig5]{ 
Maximum neutron exposure in the pocket as a function of the He intershell
\cdb from the simulations performed at the 10$^{\rm th}$ 
TDU episode of the 3 \msb model of solar metallicity computed with the MSSSP code.
The full hexagon indicate the true $X_{\rm C12}^{intershell}$ value at such
interpulse resulting with MSSSP. 
\label{fig5}}

\figcaption[fig6]{ 
Abundance profiles in the \ctb pocket found self-consistently by the
EVOL code at the 5$^{\rm th}$ interpulse period with the overshoot parameter   
$f =$ 0.128. The large tick-marks on the x-axis measure 10$^{-5}$ \msb intervals. 
Also shown is the resulting $\tau$ at the end of the interpulse as subsequently 
computed by representing the abundance profiles 
with ten zones and performing one-zone computations with
Monash nucleosynthesis code. 
\label{fig6}}

\figcaption[fig7]{ 
{\bf a)} Temperature from the three evolutionary codes, and {\bf b)} neutron density as computed by the
$s$-process code in the zone 3 of the \ctb pocket with the higher 
neutron density. The zero points in time represent
the time from the start of the interpulse period, about 10,000 yr, when in the MSSSP case $T = 0.79 \times
10^8$ K. Matter densities also grow with time from $\simeq 2\times 10^3$ g cm$^{-3}$ to $10^4$ g cm$^{-3}$
for
FRANEC and EVOL, and to $\simeq 1.5\times 10^4$ g cm$^{-3}$ for MSSSP.
\label{fig7}}

\figcaption[fig8]{ 
Primary contribution to the \nqb in the H-burning ashes as a function of thermal pulse number. Primary \nqb comes from the dredge up
of \cdb (and $^{16}$O in the EVOL case) in the envelope followed by H-shell burning. 
To compute the total mass fraction of \nqb engulfed by
each PDCZ these values have to be added to the constant value
of 0.0128 adopted in the $s$-process code for the H-burning ashes of solar metallicity.
\label{fig8}}

\figcaption[fig9]{ 
{\bf a)} The temperatures at the bottom of the PDCZ and {\bf b)} the
neutron densities as functions of time during the last thermal pulse computed in the three cases: the 
25$^{\rm th}$ with TDU for FRANEC (solid lines), the 19$^{\rm th}$ with TDU for 
MSSSP (long-dashed lines), and the 13$^{\rm th}$ with TDU for the EVOL code
(short-dashed lines). The zero in time corresponds to the instant in which
the bottom temperature in the convective shell reaches 2.5$\times$10$^8$
K. Bottom densities decrease exponentially with time from $\sim$10$^4$ to
$\sim$10$^3$ g cm$^{-3}$. The variation of temperature, matter density and 
neutron density with pulse number for the FRANEC case can be found in Figs. 
10, 11 and 13 of \citet{gallinoab}.
\label{fig9}}

\figcaption[fig10]{ 
Distribution of production factors of $s$-only isotopes (open symbols) in the He intershell material averaged
over all TDU episodes and weighed by the mass dredged up (see text). 
The production factors of the neutron-rich isotopes $^{86}$Kr, $^{87}$Rb and $^{96}$Zr, which are affected by
branchings, are also shown as full symbols. The vertical lines are eye guides drawn at the stable neutron magic
nuclei on the \s-process path: $^{88}$Sr, $^{138}$Ba and $^{208}$Pb, whose production factors
are also plotted using open dash-lined symbols. 
The production factor of the s-only isotope $^{152}$Gd is
indicated by an arrow (see text for discussion).
\label{fig10}}

\clearpage

\begin{figure}  
\plotone{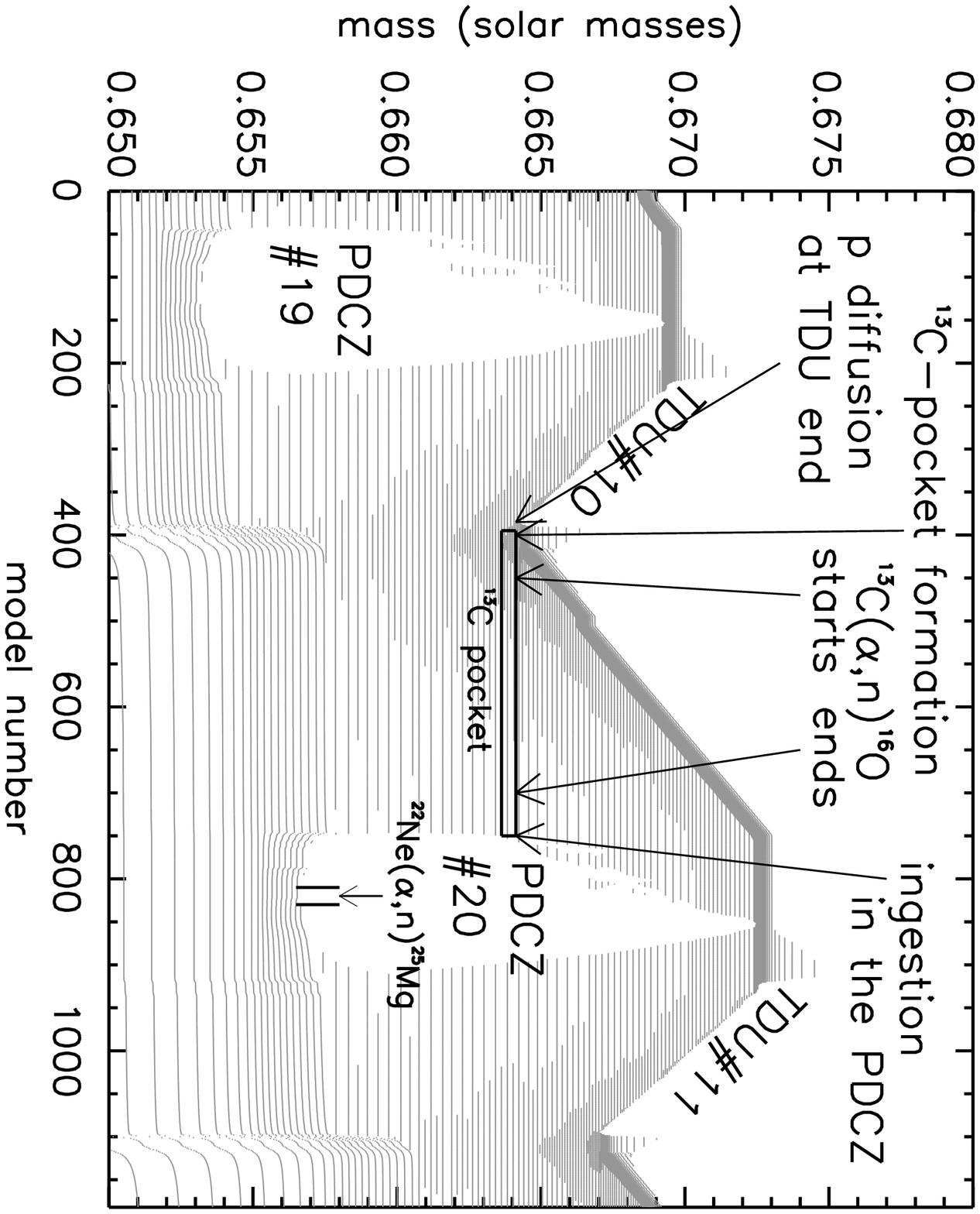}
\end{figure}

\clearpage

\begin{figure}  
\plotone{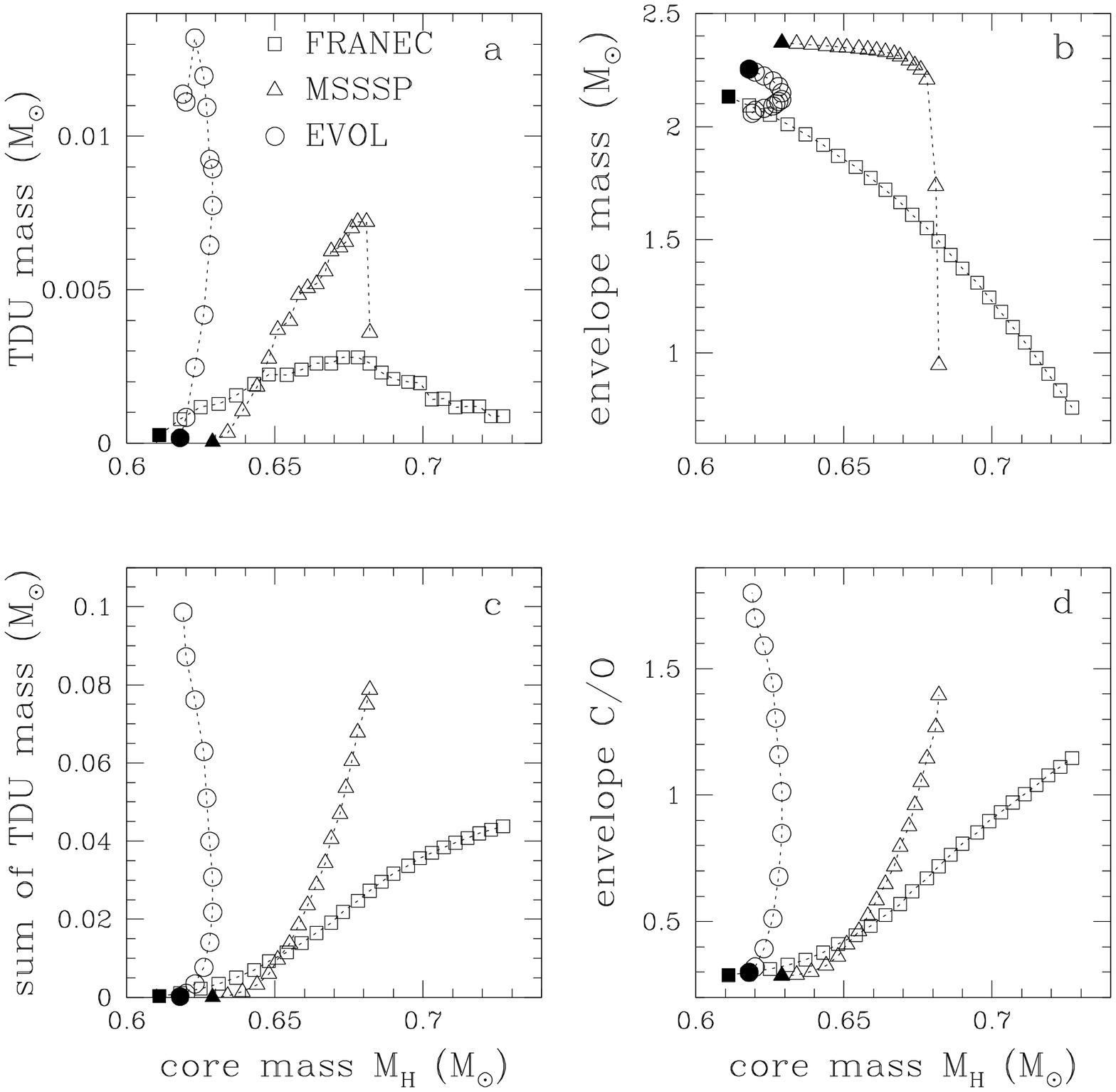}
\end{figure}

\clearpage

\begin{figure}  
\plotone{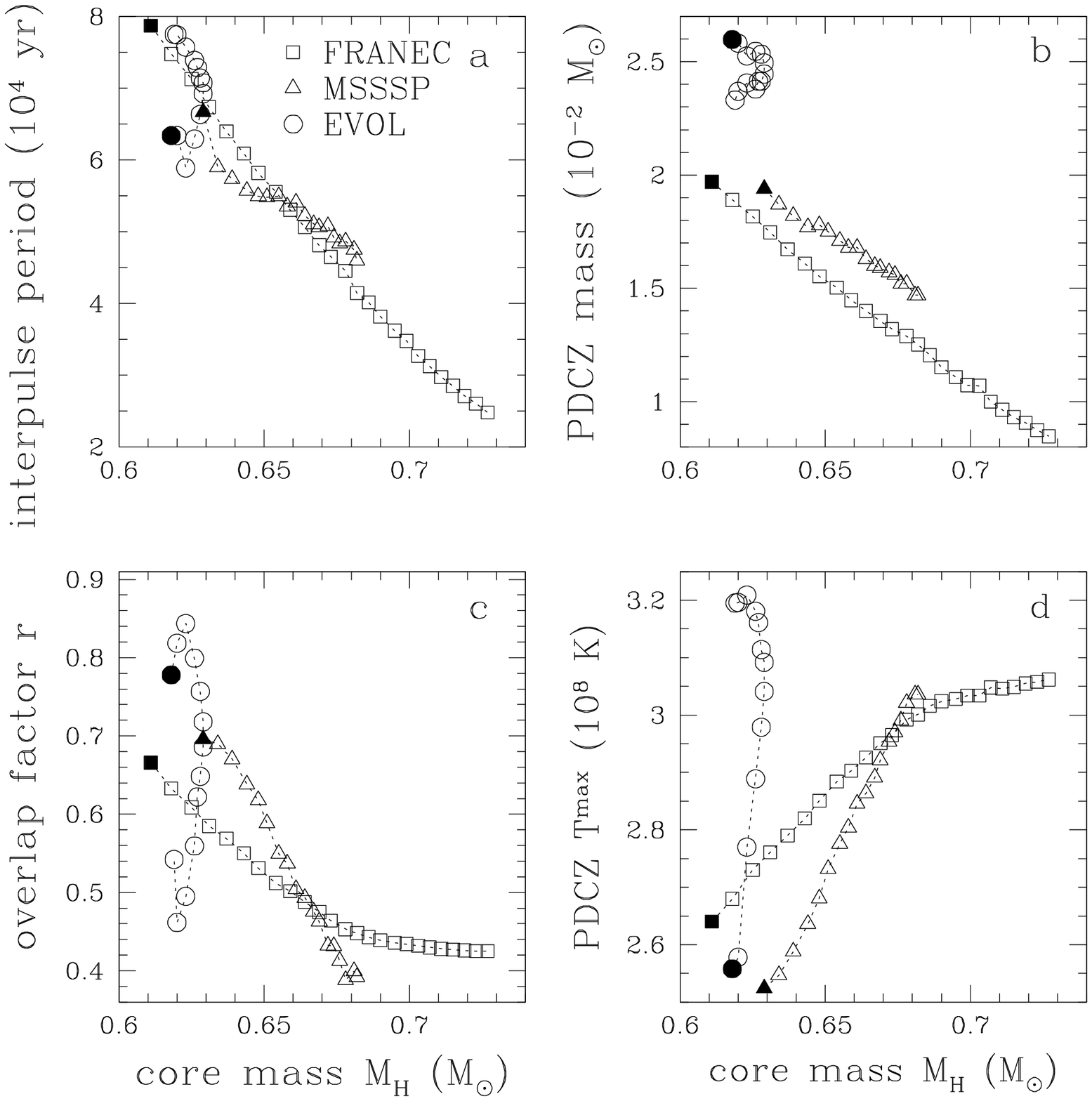}
\end{figure}

\clearpage

\begin{figure}  
\plotone{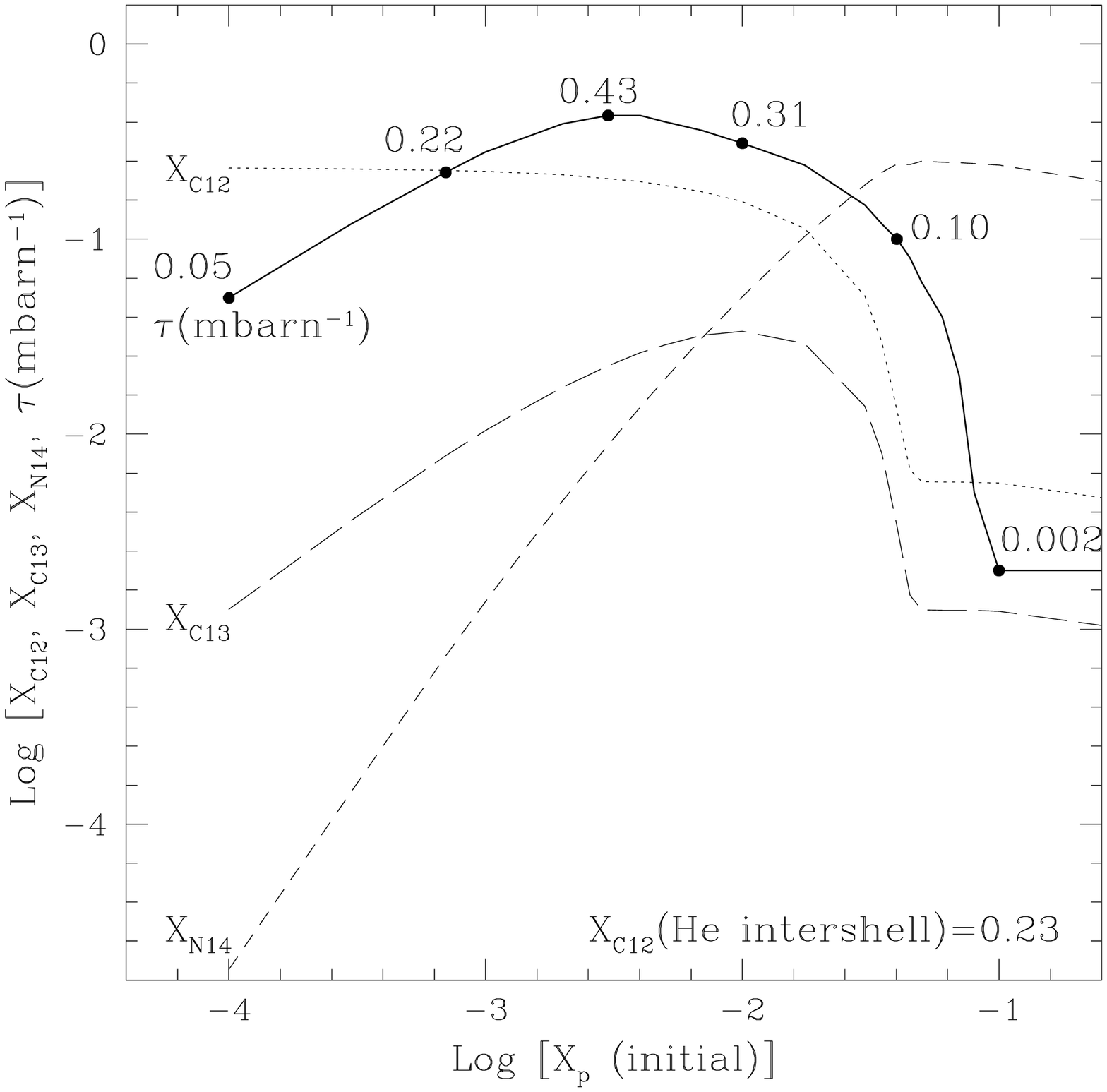}
\end{figure}

\clearpage

\begin{figure}  
\plotone{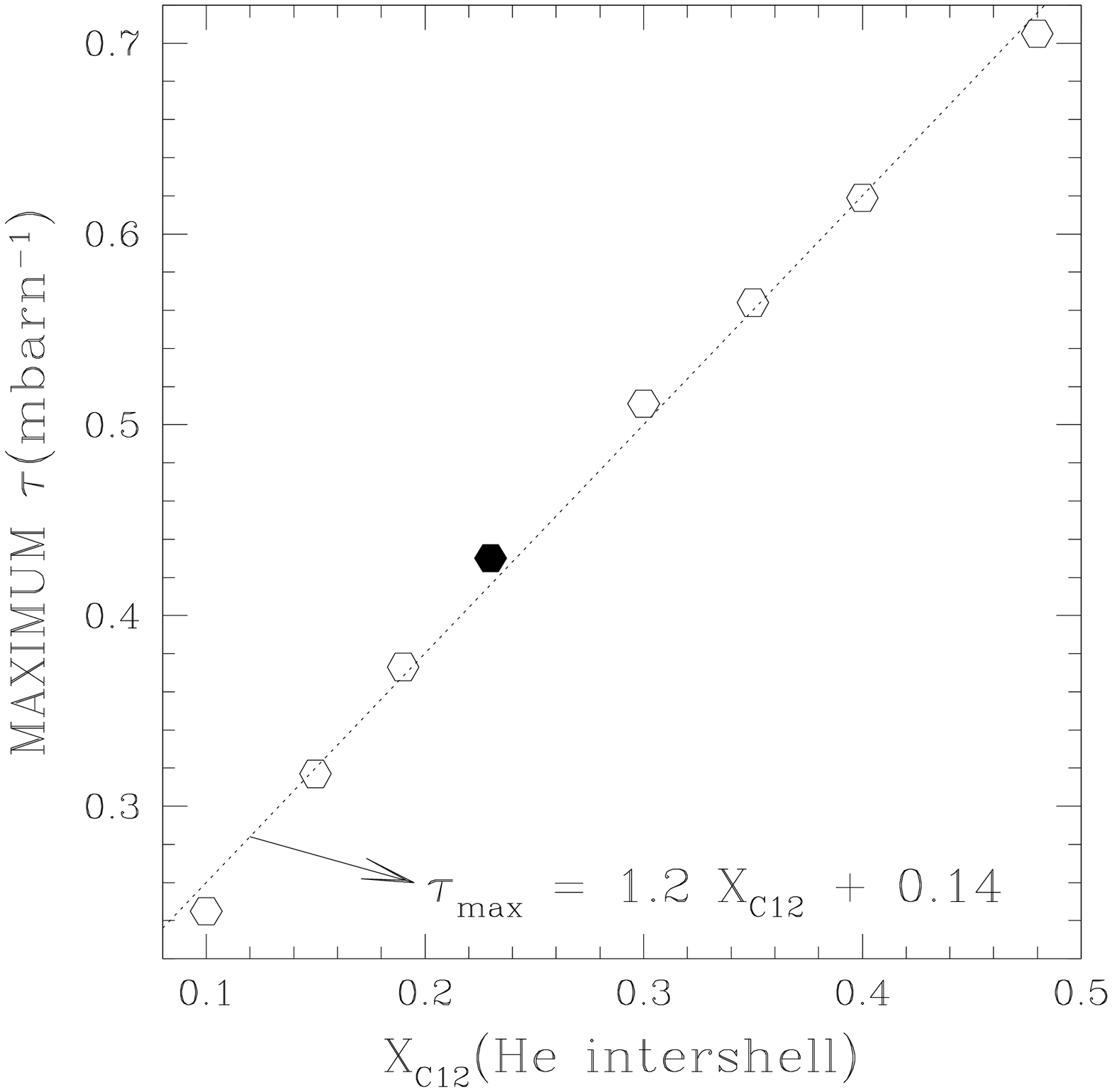}
\end{figure}

\clearpage

\begin{figure}  
\plotone{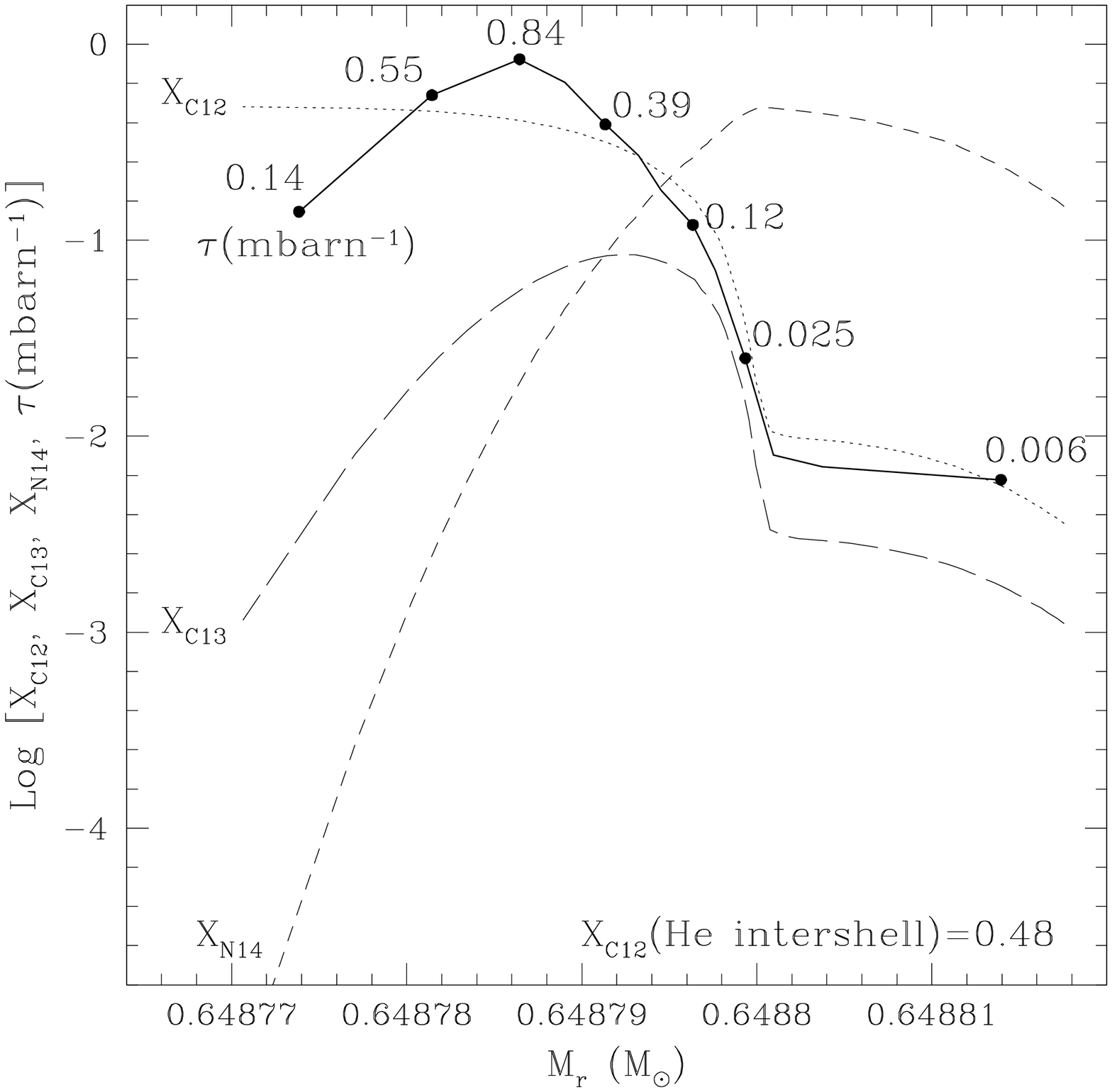}
\end{figure}

\clearpage

\begin{figure}  
\plotone{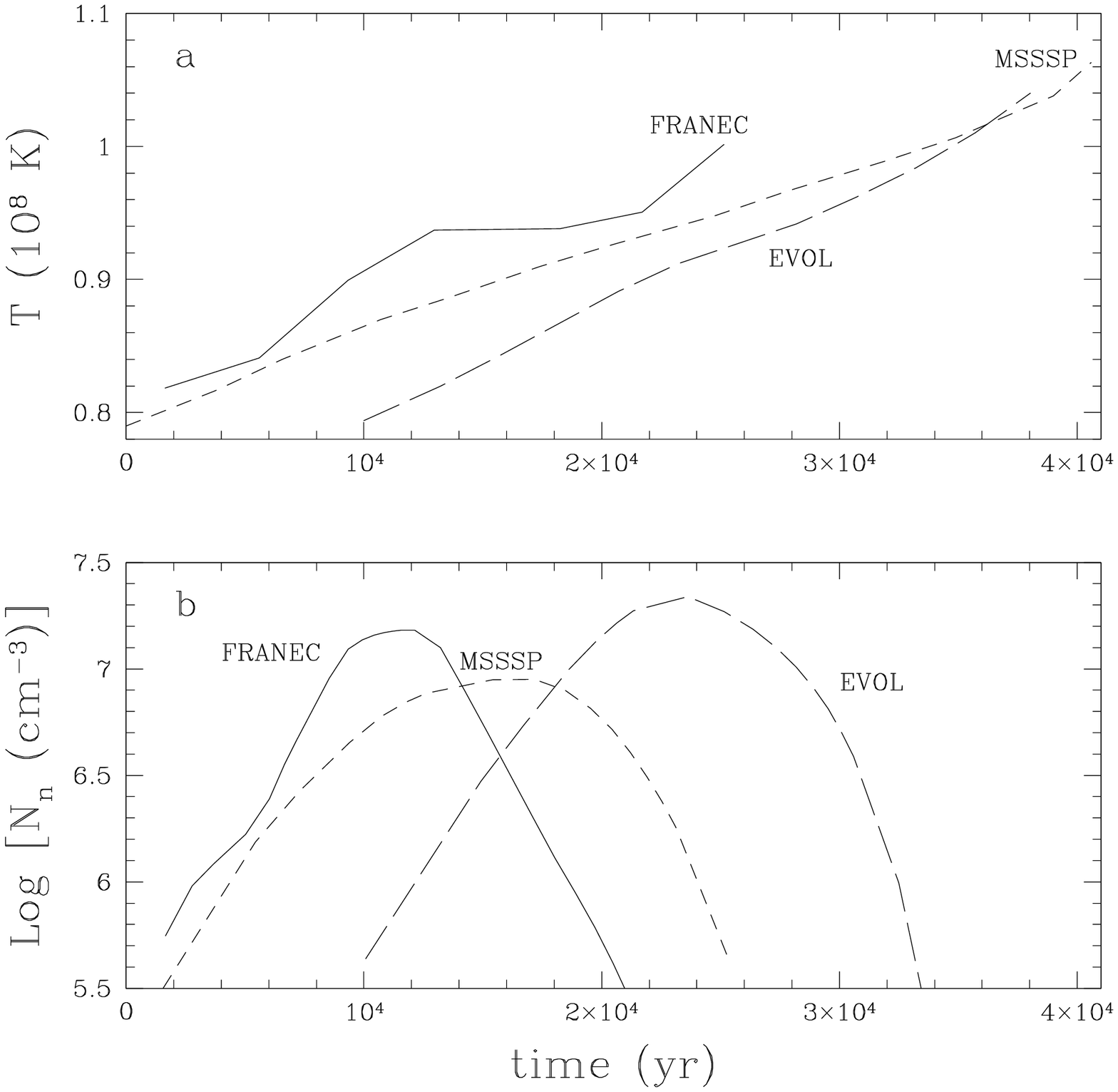}
\end{figure}

\clearpage

\begin{figure}  
\plotone{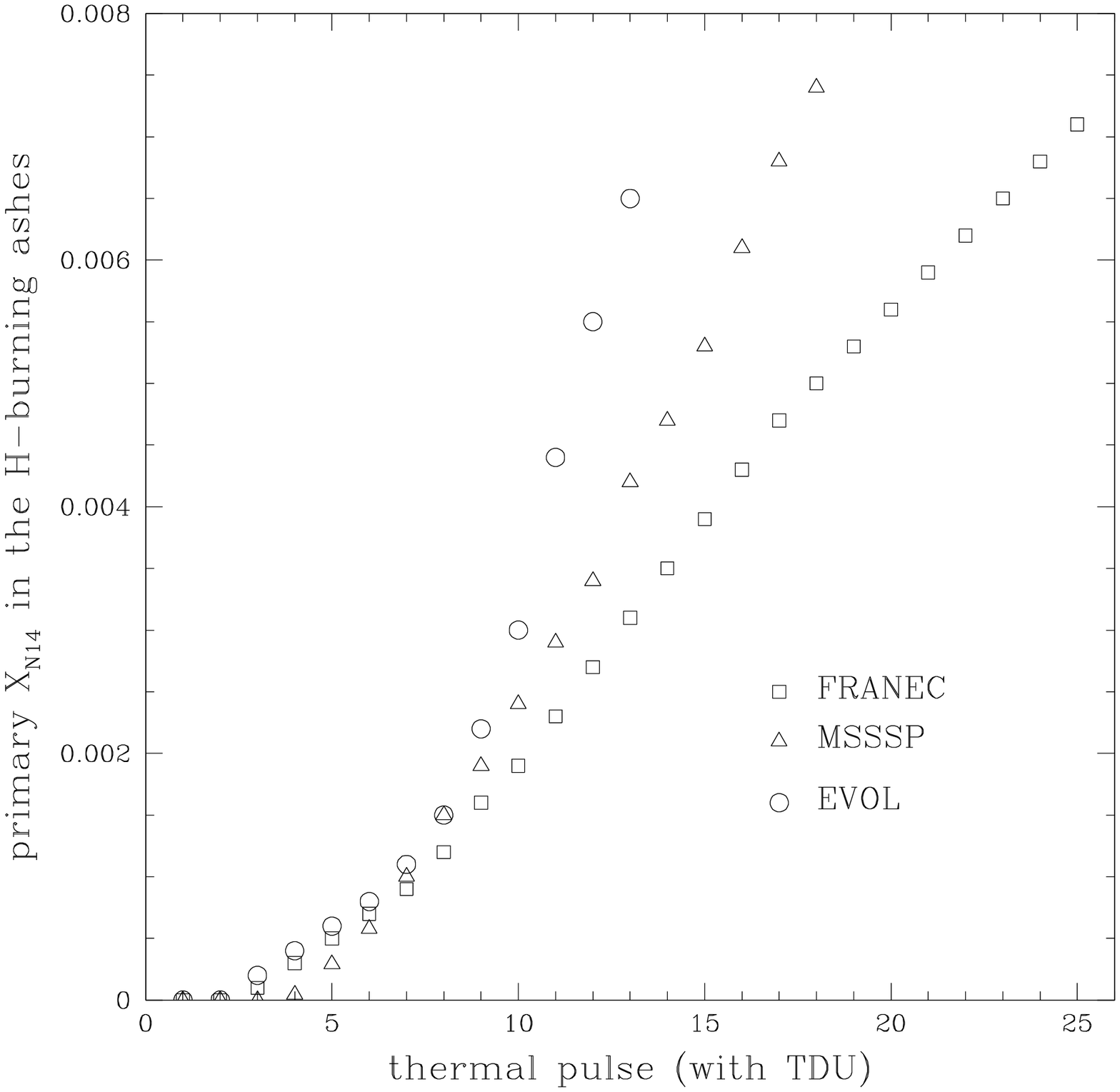}
\end{figure}

\clearpage

\begin{figure}  
\plotone{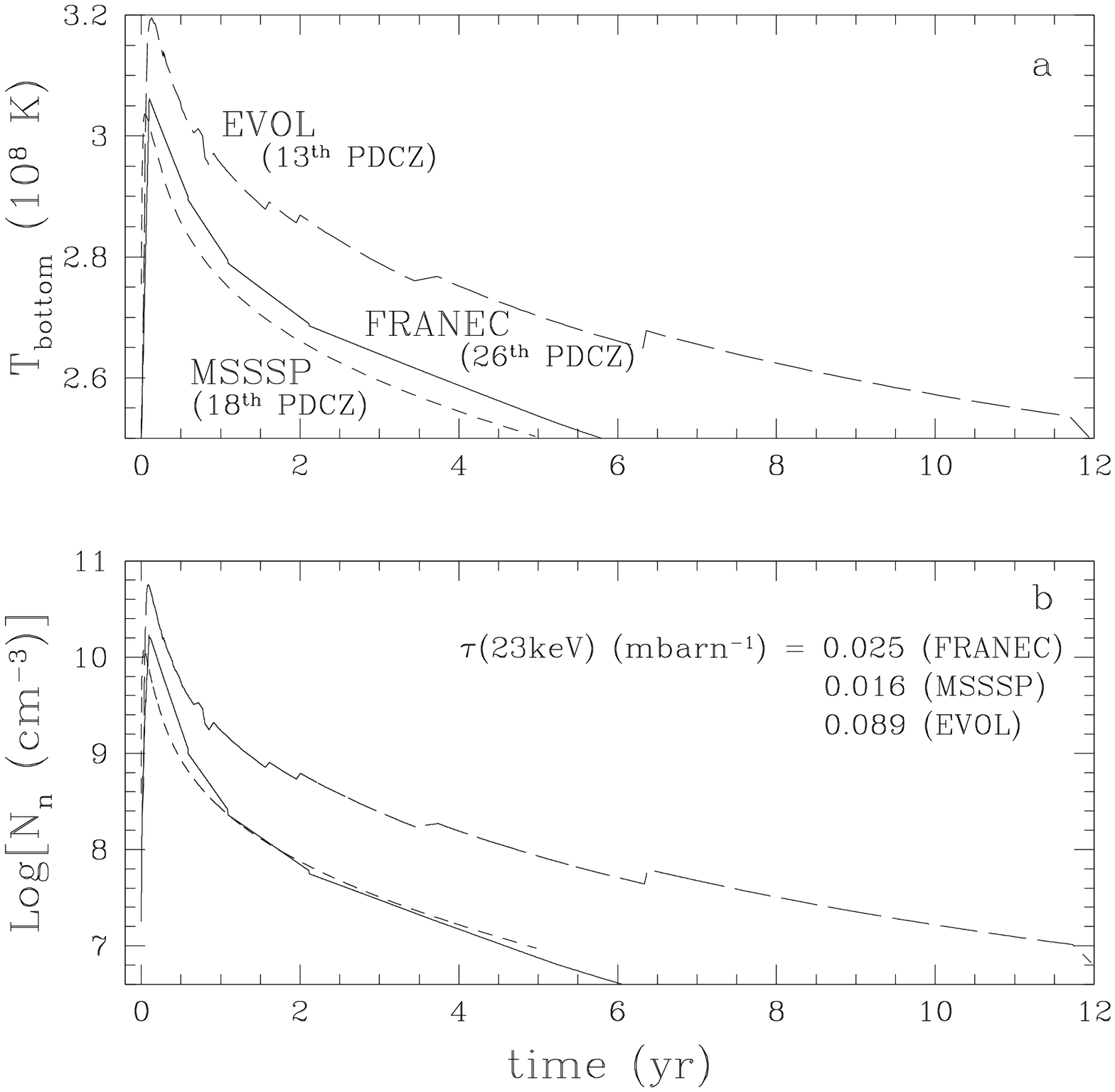}
\end{figure}

\clearpage

\begin{figure}  
\plotone{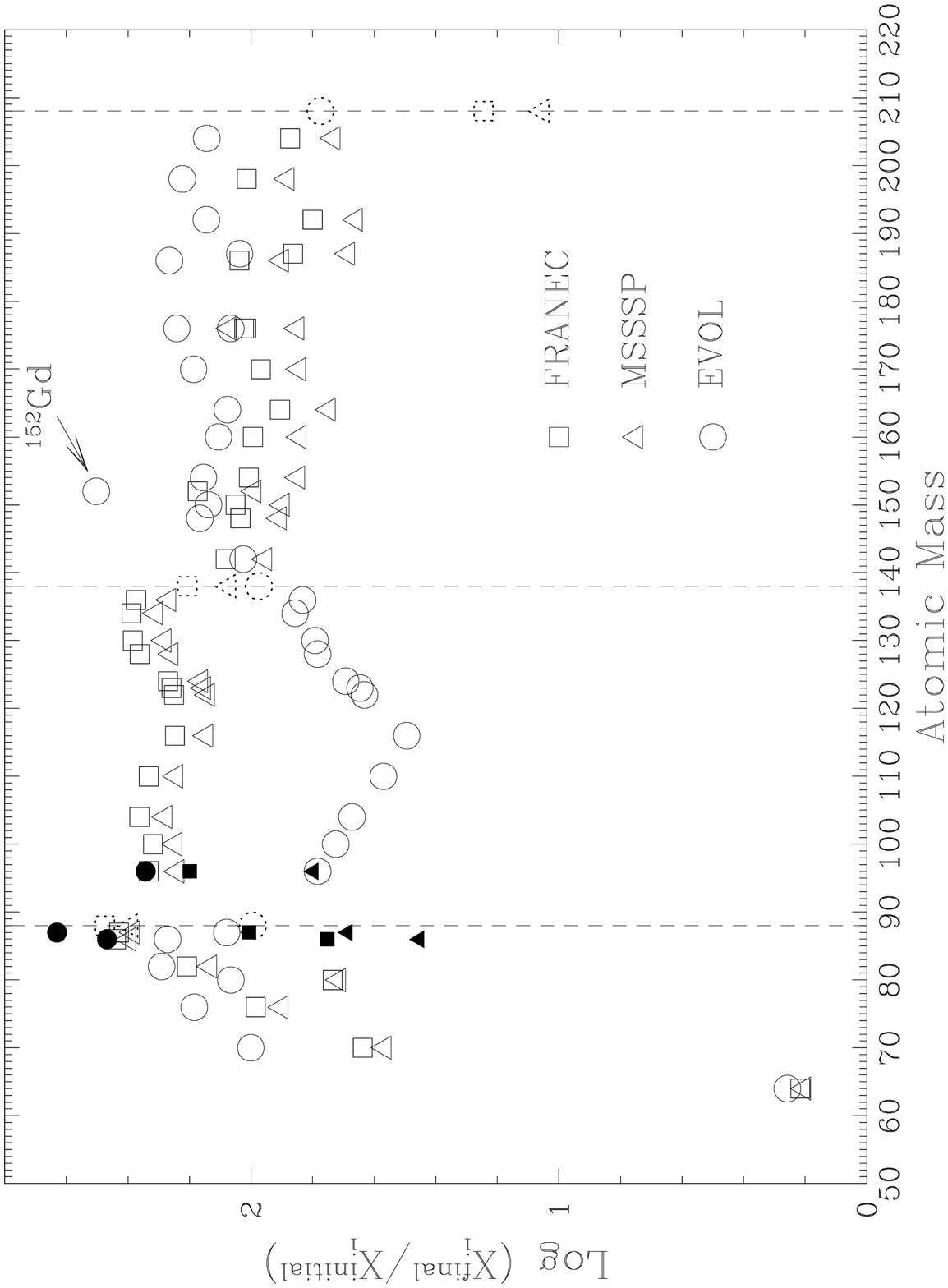}
\end{figure}

\end{document}